\documentclass[useAMS,usegraphicx]{mn2e}
\usepackage{amsmath}

\newcommand{\AAA}[3]    {\mbox{#3, A\&A,~#1,~#2}}

\newcommand{\ApJ}[3]    {\mbox{#3, ApJ,~#1,~#2}}

\newcommand{\AJ}[3]     {\mbox{#3, Astron.~J.,~#1,~#2}}
\newcommand{\MNRAS}[3]  {\mbox{#3, MNRAS,~#1,~#2}}
\newcommand{\Nature}[3] {\mbox{#3, Nature,~#1,~#2}}
\newcommand{\NewA}[3]   {\mbox{#3, NewA,~#1,~#2}}

\newcommand{\PhRevL}[3] {\mbox{#3, Phys.~Rev.~Lett.,~#1,~#2}}
\newcommand{\PhRevD}[3] {\mbox{#3, Phys.~Rev.~D#1,#2}}

\newcommand{\astroph}[1]{\mbox{preprint~(astro-ph/#1)}}

\def\ap3m{AP$^3$M}

\def\h0{$h_0$}
\def\H0{$H_0$}

\def\Msun{${\rm M}_{\odot}$}

\def\d{{\rm ~d}}

\def\ea{et~al.~}

\def\lesssim{\mathrel{\hbox{\rlap{\hbox{\lower4pt\hbox{$\sim$}}}\hbox{$<$}}}}
\def\grtsim{\mathrel{\hbox{\rlap{\hbox{\lower4pt\hbox{$\sim$}}}\hbox{$>$}}}}

\newif\ifpdf\ifx\pdfoutput\undefined\pdffalse\else\pdfoutput=1\pdftrue\fi
\newcommand{\pdfgraphics}{\ifpdf\DeclareGraphicsExtensions{.pdf,.jpg}\else\fi}

\newcommand{\bh}    {\bullet}
\newcommand{\BH}    {\mbox{\boldmath$ \bullet$}}

\begin{document}
\pdfgraphics

\title[MBH remnants of the first stars I: abundance
in present-day galactic haloes]{Massive black hole remnants of the
  first stars I: abundance
in present-day galactic haloes}

\author[Islam R.R., Taylor J.E. \& Silk J.]
       {Ranty R. Islam\footnote{Email: rri@astro.ox.ac.uk}, James E. Taylor and Joseph Silk\\
       {Astrophysics, Denys Wilkinson Building, Keble Road, Oxford, OX1 3RH, UK}}

\date{Received ...; accepted ...}

\maketitle

\begin{abstract}
We investigate the possibility that present-day
galaxies and their dark matter haloes contain
a population of massive black holes (MBHs) that form by
hierarchical merging of the black hole remnants of the first
stars in the Universe.
Some of the MBHs may be large enough or close enough to the
centre of the galactic host that they merge within a Hubble time. We
estimate to what extent this process could contribute to the mass
of the super-massive black holes (SMBHs) observed in galactic
centres today. The relation between SMBH and galactic bulge mass
in our model displays the same slope as that found in observations.
Many MBHs will not reach the centre of the host halo,
however, but continue to orbit within it. In doing so MBHs may
remain associated with remnants of the satellite halo systems 
of which they were
previously a part.
Using a semi-analytical approach that explicitly accounts for dynamical
friction, tidal disruption and encounters with galactic disks, we
follow the hierarchical merging of MBH systems and their subsequent
dynamical evolution inside the respective host haloes.
In this context two types of dynamical process are examined in
more detail.
We predict the mass and abundance of MBHs in present-day
galactic haloes, and also estimate the MBH mass accretion rates as well
as bolometric luminosities for two different accretion scenarios.
MBHs that have not undergone recent merging will retain associated  dark matter
cusps that were enhanced  by  black hole accretion growth,
and may be possible sources of gamma rays via neutralino annihilations.
\end{abstract}

\begin{keywords}
galaxies: formation -- galaxies: haloes -- galaxies: nuclei --
cosmology: theory
\end{keywords}

\section{Introduction}
The presence of super-massive black holes (SMBHs) at the centres of most
galaxies appears by now to be firmly established. SMBHs have estimated
masses in the range $10^6 - 10^9$ ~\Msun and a number of correlations
have been observed between the mass of 
SMBHs and properties of the galactic bulge hosting them.
The first of these to be established were correlations between the
mass of the SMBH, $M_{smbh}$ and the mass or luminosity of the
galactic bulge,  $M_{bulge}$ and $L_{bulge}$ 
respectively \cite{magorrian98,kormendy01,laor01}.
More recently, a tighter correlation was found between $M_{smbh}$ and
the bulge velocity dispersion, $\sigma_{bulge}$
\cite{gebhardt00,merritt01}, and also between 
$M_{smbh}$ and the bulge's light profile, as parameterised by a shape
index, $n$ \cite{graham01}.

Since these correlations extend well beyond the direct dynamical
influence  of the SMBH it seems likely that there is a close link
between the formation of SMBHs and the formation of their host galaxy.
A recent analysis finds that the masses of SMBHs appear to be
correlated with the host circular velocity even beyond the optical
radius \cite{ferrarese02}. If confirmed, this implies that the SMBHs are linked
to properties of the host dark matter halo. This would be the
strongest hint yet that there must be a hierarchical merging component to
the growth of SMBHs, since the properties of haloes are primarily
determined in the context of their hierarchical build up.

Most models put
forward to account for the correlations assume a close link between
galaxy and SMBH formation as a starting point. We can distinguish two
generic types of models.
One proposes that the SMBH mass increases mainly by the merging
of smaller precursors. 
This requires SMBH precursors to have been present in galaxies from very
early on \cite{madau01,menou01,schneider02} 
It might allow the observed correlations to be set up over a long period 
of time with a potentially large number of mergers through the
dynamical interactions between the merging galaxies and SMBH
precursors.
 
Another mechanism considered is growth mainly by gas accretion within the 
host bulge. In this case a strong non-gravitational interaction between the
growing SMBH and the bulge is required. An example of this is
the radiative feedback of an accreting SMBH that changes the gas dynamics
in the bulge so as to effectively control its own gas supply and establish a
relation between $M_{smbh}$ and $\sigma_{bulge}$ \cite{silk98}.
A similar route is followed by  models that tie $M_{smbh}$ to the amount and properties of gas
in the bulge \cite{adams01}.
A combination of both approaches is used in the model by Haehnelt \&
Kauffmann (2000) \nocite{haehnelt00}.

As an example of the merger-only scenario it has been shown that the
merging of the massive black hole (MBH) remnants of the first stars in
the Universe could account for the inferred overall abundance of SMBHs
today \cite{schneider02}.
 
Here we explore this idea further to determine an upper limit
on the mass to which SMBHs can grow through mergers of
lower mass precursors and more importantly what the implications are for the presence of a remnant 
population of lower mass MBHs in the galactic halo. In doing so we
assume efficient merging between MBHs, but we also consider the effect
of relaxing this assumption.
As the `seeds' in the merging hierarchy, we consider massive
black holes (MBHs) of some mass $M_{seed}$ that are remnants of the
first stars in the Universe, forming within high density
peaks at redshifts of $z \sim 20 - 30$. We use
Monte Carlo merger trees to describe the merging of haloes and then
follow the dynamical evolution of merged/accreted satellite haloes and
their central MBHs within larger hosts, explicitly accounting for dynamical
friction, tidal stripping and disk encounters. 
A key prediction is that $\sim 10^3$ MBHs in the mass range $1 -
1000 \times M_{seed}$  should be present within the galactic halo today as a
result of this process.

In a previous paper \cite{islam03} we looked at the case of 260 \Msun
~seed MBHs forming within 3$\sigma$ peaks at redshift $z \sim 24$.
In this paper we extend our investigation to consider four different sets of initial conditions as
shown in table \ref{tab:inits} and also compute the accretion rates of
resulting MBHs in present-day haloes.
The latter is used to determine the bolometric accretion
luminosities. In a subsequent paper we use specific spectral models to
predict optical and X-ray luminosity functions for the accreting halo MBHs.
Throughout we work with a $\Lambda$CDM cosmology, specified by $\Omega_m =
0.3, \Omega_{b} = 0.02 h^{-2}, \Omega_{\Lambda} = 0.7, h = 0.7, \sigma_8 = 0.9$.

The structure of this paper is as follows. In section
\ref{sec:MBHmerging} we briefly describe how MBHs could form as a result of
the formation and evolution of the first stars in the Universe. We
also introduce the semi-analytical scheme used to track the subsequent
merging and dynamical evolution of MBHs.
In section \ref{sec:MBHpresentday} we present the resulting
distribution of MBHs in galactic haloes today. We also discuss the MBH
accretion rates and to what extent
these MBHs could have contributed to the mass of the SMBHs in galactic centres. 
In section \ref{sec:observations} we estimate the bolometric
luminosities of accreting MBHs. This should give some indication of
the expected actual observable signatures. 
A summary of our findings and conclusion is given in section \ref{sec:conclusions}.

\section{Formation and hierarchical merging of seed mass MBHs}
\label{sec:MBHmerging}
\subsection{Primordial star formation and MBHs}
A number of recent semi-analytical \cite{hutchings02,fuller00,tegmark97} and
numerical investigations \cite{bromm02,abel00} suggest that the
first stars in the Universe were likely created inside molecular
clouds that fragmented out of the first baryonic cores 
inside dark matter haloes at very high redshifts.
For common $\Lambda$CDM cosmologies in particular these {\em minihaloes} are
found to have a mass $M_{min} \sim 10^5  - 10^6 ~h^{-1}$ \Msun and to
have collapsed at redshift $z
\sim 20 - 30$. In linear collapse theory this corresponds to collapse from
$2.5 - 3.5 \sigma$ peaks in the initial matter density field.
For instance minihaloes collapse from $3 \sigma$ peaks at a
redshift of about 25.
This is because the mass
contained in overdensities corresponding to $3 \sigma$ peaks at this
redshift is just above both the cosmological Jeans mass and the 
cooling mass \footnote{At or above the cooling mass the corresponding
virial temperature, to which the baryons are heated, is high enough for
cooling to proceed on a time scale that is smaller than the gravitational infall
time scale. The latter is the necessary condition for fragmentation to occur.}.
Cooling nevertheless proceeds much more slowly than at present; as
stars have yet to form, metals that could facilitate more efficient 
cooling are essentially not present. 
This implies that even though fragmentation occurs, fragments will be
much larger than in a corresponding situation today. 
Seed masses within these 
fragments can in principle accrete large amounts of matter from the
cloud without further fragmentation occurring, which could eventually
lead to the formation of a proto-star.
\cite{bromm02,machacek01,omukai01}. Only radiation pressure from 
the proto-star on the infalling layers of material could halt
accretion and so limit the mass of the star. However, in the absence of dust the
infalling matter has too low an opacity for radiation pressure to be
significant \cite{ripamonti02}. In these stars the role of winds that 
could lead to significant mass loss in population I stars, is also negligible.
As a result this will likely lead to the creation of very massive
stars, potentially as heavy as $10^3$ \Msun. These are also referred
to as population III stars. 

As yet nothing definite is known about the initial mass function (IMF) 
of these stars. However, their large mass will see 
many of them ending up as black holes of essentially the same mass -
gravity is so strong that not even ejecta of a final supernova can
escape \cite{heger02}. 

Here we assume that in each dark matter halo forming at $z \grtsim 20$ with
a mass larger than the cooling mass, one MBH forms as the end result of any primordial star formation
occurring inside the halo.
These MBHs then represent the seeds for the subsequent merging process.

For our computations we have considered three different formation
redshifts of minihaloes and seed MBH masses as summed up in table \ref{tab:inits}.
Our choice of a seed MBH mass of 260 \Msun is motivated by the result
of Heger (2002) \nocite{heger02} that massive stars above this mass
will not experience a supernova at the end of their lives but will
collapse directly to a MBH of essentially the same mass.

\begin{table}
  \begin{center}
    \caption{Masses of seed MBHs and heights of peaks in initial density field
      within which they formed.}
    \begin{tabular}{lccc} \\	\hline \hline \label{tab:inits}
      & $M_{\bh,seed}$ & peak height $\nu_{pk}$ & $z_{collapse}$\\ \hline
      A & 260 \Msun & 3.0 & 24.6\\
      B & 1300 \Msun & 3.0&  24.6\\
      C & 260 \Msun & 2.5 & 19.8\\
      D & 260 \Msun & 3.5 & 29.4\\ \hline \hline
    \end{tabular}
  \end{center}
\end{table}

\subsection{Hierarchical merging and dynamical evolution of MBHs}
\label{sec:semianalytical}
While the basic properties of the seed MBHs are determined by the 
physics of the first baryonic objects, as outlined above, the extent
to which they merge to form the present-day SMBH depends on their 
subsequent dynamical evolution after their respective host haloes 
have merged. To track this evolution we use a semi-analytical code 
(Taylor \& Babul 2001 and 2003\nocite{taylor01,taylor03}) that
combines a Monte-Carlo algorithm to generate halo merger trees with 
analytical descriptions for the main dynamical processes -- 
dynamical friction, tidal stripping, and tidal heating -- 
that determine the evolution of merged remnants within a galaxy halo.

Starting with a halo of a specific mass at the present-day, we trace
the merger history of the system back to a redshift of 30, using
the algorithm of Somerville \& Kolatt (1999). Computational 
considerations limit the mass resolution of the tree to 
$\simeq\,3\times\,10^{-5}$ of the total mass; below this limit 
we do not trace the merger history fully. For the more massive
haloes, this resolution limit is larger than $M_{min}$ and many 
of the branchings of the merger tree drop below the mass resolution 
limit before they reach the minihalo collapse redshift (e.g. $z = 24$
for collapse from $3 \sigma$ peaks), so that we cannot always track the 
formation of individual black holes. To overcome this problem, if 
systems over $M_{min}$ appear in the merger tree after primordial 
black holes have started forming at the collapse redshift, we determine how
likely they are to contain one or more primordial black holes, 
based on the frequency of peaks of corresponding height, and populate them accordingly. 
In the most massive trees, haloes at the resolution limit are likely
to contain several primordial black holes. In this case, we assume the 
black holes have merged to form a single object, in keeping with
the assumption of efficient merging discussed below.

Within the merger trees, we then follow the dynamical evolution of
black holes forward in time to the present-day, using the analytic model
of satellite dynamics developed in Taylor \& Babul (2001). 
Merging subhaloes are placed on realistic orbits at the virial radius of 
the main system, and experience dynamical friction, mass loss and heating
as they move through their orbits. The background potential is modelled
by a smooth Moore profile, $\rho \propto r^{-1.5}(r_s^{-1.5} + 
r^{-1.5})$ \cite{moore99}, which grows in mass according to 
its merger history, and changes in concentration following the relations 
proposed by Eke, Navarro \& Steinmetz (2001). We give this profile a 
constant-density core
of radius $0.1 r_s$, to account for the possible effects of galaxy 
formation in disrupting the dense central cusp. 

Within this potential, the formation of a central 
galaxy with a disk and a spheroidal component is modelled schematically,
by assuming that a third of the gas within the halo cools on the dynamical
time-scale to form a galactic disk, and that major mergers disrupt this 
disk and transform it into a spheroid with some overall efficiency. 
We choose as the disruption criterion that the disk collide with
an infalling satellite of mass equal to or greater than its own, 
and set the efficiency with which disk material is then transferred 
to the spheroid to 0.25. This choice of parameters is required to
limit the formation of spheroids and thus produce a reasonable range 
of morphologies in isolated present-day $10^{12} M_{\odot}$ systems, 
as discussed in Taylor \& Babul (2002). We do not expect the results 
for halo back holes to depend strongly on these parameters, although
they may have some effect on the properties of the central black holes. 
Finally, the evolution of 
haloes in side branches of the merger tree is followed more approximately,
by assuming that higher-order substructure (that is subhaloes within subhaloes)
merges over a few dynamical times, causing its black hole component to 
merge as well, while unmerged substructure percolates down to a
lower level in the tree. We will discuss the details
of this model elsewhere (Taylor \& Babul 2003 \nocite{taylor03});
here it serves only as a backdrop for the dynamical calculations of
black hole evolution. 

The semi-analytic code tracks the positions of all the primordial black holes 
that merge with the main system and the amount of residual dark matter 
from their original halo that still surrounds them, if any. 
We classify systems as `naked' if their surrounding 
subhalo has been completely stripped by tidal forces, and `normal'
otherwise. Our orbital calculations cannot follow the evolution of
systems down to arbitrarily small radii within the main potential, so if 
black holes come within
1\% of the virial radius of the centre of the potential (roughly 3 kpc
for a system like the present-day Milk Way), we assume they have `fallen in'
and stop tracking their orbits. Black holes contained in satellites
which disrupt the disk in major mergers are also assumed to fall into
the centre of the potential during its subsequent rearrangement. 
Clearly, this assumes that black hole merging in the centre of the main 
system is completely efficient, so it will produce a conservative upper 
limit on how many black holes merge with the central SMBH. We discuss
the effect of relaxing these assumptions below. 

Using the semi-analytic code, we generate sets of different realisations 
for final halo masses of $1.6\times10^{10}$, $1.6\times10^{11}$, 
$1.6\times10^{12}$ and $1.6\times10^{13}$ \Msun for the four sets of
initial conditions given in table \ref{tab:inits}.

\subsection{MBH merger efficiency}
\label{sec:mergeff}
Up to now we have considered any MBH as having merged with the central 
MBH, when it comes within  one per cent (hereafter referred to as the
merger region) of the virial radius of the
host halo at that time.
There are various ways in which the actual merger efficiency could be
lower than this, and so our results above only provide an upper limit
on how much the MBH merger process can contribute to the mass of
central and halo MBHs.

One major source of inefficiency is of course the time it takes for
any MBH to spiral into another, and typically more massive MBH at the
centre of their common host and how likely it is then for the two to
merge. One does not necessarily imply the other -- at early
times haloes are smaller, that is, at the first encounter, 
the two central MBHs within any two haloes will start out much
closer and so are more likely to spiral to the common centre of the
halo merger remnant in a relatively short time.
Because there are more low mass haloes, this might then give rise to configurations consisting of 
more than two MBHs and thus the possibility of sling-shot
ejections. In other words some fraction of MBHs, although having
travelled to the centre quickly, might eventually end up being expelled 
rather then merging. 
 This has implications for the most
massive trees. Haloes at the resolution limit in these trees have a
mass above $M_{min}$ and therefore might appear in the tree with
several seed MBHs which we have thus far assumed to have merged to form
one MBH (c.f. section 3.2). This may no longer be the case if
slingshot ejections occur. Assuming that in this case the lightest
MBHs are ejected, however, this should not significantly reduce the
mass of the central MBH.

The assumption that MBHs within a kpc or so from the host centre merge 
efficiently can be used to determine an upper limit on the mass of
central SMBHs. Although MBH merging may proceed much less
efficiently, a number of processes could lead to
rapid merging of MBHs in the galactic context. 

If the mass of only the MBHs is considered, their orbital decay time
scale in the host can be longer than a Hubble time.
However, MBHs typically remain associated with matter from
their original satellite, which increases their effective mass by a
factor of at least 100 to 1000 and lowers the orbital decay time scale
accordingly, allowing even relatively light MBHs ($M_{\bh} \ge 10^3$
\Msun) to spiral into the host central
region ($\lesssim$ kpc) within a Hubble time \cite{yu02}. This is true even if the
satellite itself may have actually lost most of its mass ($\grtsim$ 99 percent)
due to tidal stripping inside the host halo and is thus classified as
`naked' in our treatment.
This implies that only at high redshifts could seed mass MBHs have travelled to the host 
centre, since they would have then entered the correspondingly
smaller host halo at smaller distances from the centre.

It seems then that dynamical friction can deliver MBHs to the host central
regions efficiently where they then form binaries with any MBH already 
at the centre. The evolution of a MBH binary system in stellar
background has been studied
extensively \cite{begelman80,quinlan96,milos01,yu02} and the
`hardening' stage of binary evolution has been
singled out as the `bottle neck' on the way to the final merger
\cite{milos01,yu02}. With dynamical friction no longer significant
and orbital decay due to gravitational wave emission not yet
important, one way for the binary MBHs to reduce their orbital
radius is
by interaction with stars in their vicinity, which can take
significantly longer than a Hubble time.
However, Merritt \& Poon (2003) \nocite{merritt03} showed recently that if the galactic
potential around a binary SMBH at its centre is non-axisymmetric and
the stellar orbits are chaotic, the interaction rates can be larger by
orders of magnitude. In principle this argument should also hold for
MBHs in their respective satellites. In fact, due to the higher number
of major mergers at early times the potentials of the (mini) galaxies
hosting central MBHs are much more likely to be non-axisymmetric.

However, even in the case that binary decay through interactions with stars
takes prohibitively long, the presence of gas may be of crucial
importance in this context \cite{milos01}. High densities of gas between the binary
MBHs could allow for a much faster evolution and eventual 
merger of the binary. Several scenarios have been suggested for this,
such as a massive gas disk around the binary \cite{gould00,armitage02} or massive 
gas inflow (see e.g. Begelman, Blandford \& Rees 1980) in the wake of major mergers.
Hydrodynamical simulations of galaxy mergers, for instance,  find that up to 60 per cent of the total gas mass of two
merging Milky Way-size galaxies can end up within a region only a few hundred
parsecs across, which is about half the bulge scale radius
\cite{barnes96,naab01,barnes02}. 
 
Here we assume that during major mergers the gas infall will actually
lead to all MBHs binaries merging. We also neglect the possibility of
triple BH interactions and sling-shot ejections.

\section{MBHs in present-day galactic haloes}
\label{sec:MBHpresentday}
A fundamental prediction of the hierarchical merging of MBHs is the existence of MBHs throughout
galaxies and their haloes. Other MBHs may have already travelled to the centre and merged there to help build up
the central SMBH we see in galactic centres today. In this section we
determine the number, mass, accretion rates and bolometric accretion
luminosities for halo MBHs and the mass of the central (S)MBH.
\subsection{Abundance and Mass of MBHs in galactic haloes} \label{sec:MBHabundance}
\subsubsection{Abundance of MBHs in galactic haloes} 
\begin{figure}
  \centering
  \includegraphics[width=84mm]{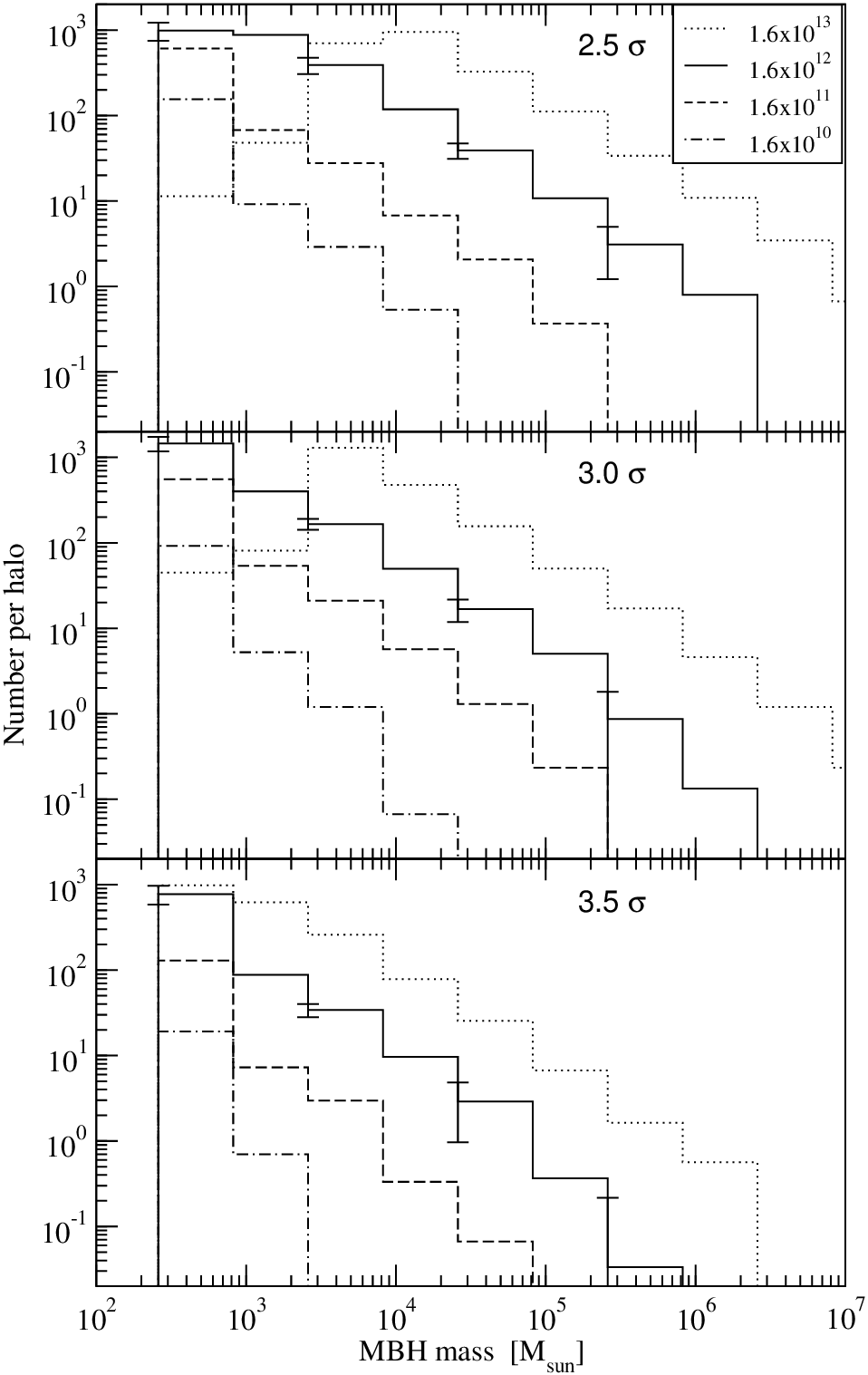}
  \caption{Abundance of all MBHs in the halo for models C, A and D (top to
    bottom panels) averaged over 30 trees with error bars corresponding
    to the standard deviation.}
  \label{fig:MBHmassfunc}
\end{figure}

In figure
\ref{fig:MBHmassfunc} we show the abundance of all
MBHs for models A, C and D and for all final halo masses. Model B,
which only differs from model A by its
different MBH seed mass is considered below.

In addition the total number of MBHs in the halo is given in table
\ref{tab:global_abund}. For Milky Way sized
haloes (i.e. corresponding to a final halo mass of $1.6 \times 10^{12}$
\Msun), for instance, we would expect between 900 to 2100 MBHs to
orbit within the galactic halo depending on the
model.

\begin{table*}
  \centering
  \begin{minipage}{100mm}
    \caption{Number of MBHs in halo averaged over thirty trees with
      associated standard deviation. We have shown the total number as
      well as the number in the bulge within two
      bulge scale radii and in the disk within two disk scale lengths and
      scale heights.}
    \label{tab:global_abund}
    \begin{tabular}{lllll} \\ \hline \hline
      Halo mass & model A & model B & model C & model D \\ \hline
      \multicolumn{5}{c}{total \# MBHs in halo} \\ \hline
      $1.6\times10^{10}$ & 100 $\pm$ 17 & 76 $\pm$ 13      & 170 $\pm$ 38 & 20 $\pm$ 6\\
      $1.6\times10^{11}$ & 640 $\pm$ 110 & 570 $\pm$ 120   & 710 $\pm$ 76 & 140 $\pm$ 24\\
      $1.6\times10^{12}$ & 2090 $\pm$ 360 & 1750 $\pm$ 170 & 2430 $\pm$ 550 & 910 $\pm$ 220\\
      $1.6\times10^{13}$ & 2130 $\pm$ 230 & 2250 $\pm$ 490 & 2200 $\pm$ 530 & 1970 $\pm$ 210\\
      \hline
      \multicolumn{5}{c}{\# MBHs in bulge} \\ \hline
      $1.6\times10^{10}$ & 0              & 0              & 0              & 0 \\
      $1.6\times10^{11}$ & 2.3 $\pm$ 1.7  & 2.3 $\pm$ 1.9  & 4.1 $\pm$ 2.6  & 0.5 $\pm$ 0.7\\
      $1.6\times10^{12}$ & 36 $\pm$ 17    & 35 $\pm$ 13    & 36 $\pm$ 11    & 11 $\pm$ 6\\
      $1.6\times10^{13}$ & 93 $\pm$ 30    & 72 $\pm$ 25    & 77 $\pm$ 26    & 81 $\pm$ 29\\
      \hline
      \multicolumn{5}{c}{\# MBHs in disk} \\ \hline
      $1.6\times10^{10}$ & 0              & 0              & 0              & 0 \\
      $1.6\times10^{11}$ & 0.9 $\pm$ 1.1  & 1.1 $\pm$ 1.2  & 2.3 $\pm$ 1.8  & 0.3 $\pm$ 0.6\\
      $1.6\times10^{12}$ & 9.3 $\pm$ 5.2  & 9.2 $\pm$ 4    & 9.2 $\pm$ 3.5  & 3.2 $\pm$ 2\\
      $1.6\times10^{13}$ & 14.6 $\pm$ 6.6 & 10.3 $\pm$ 4.3 & 12.6 $\pm$ 4.8 & 11.7 $\pm$ 5\\  \hline\hline
    \end{tabular}
  \end{minipage}
\end{table*}
Here, as in all other plots of average abundances and luminosities, we have stated the standard
deviation rather than the error on the `mean' as a measure of the
uncertainty of our results.

We found that the number of MBHs in the galactic disk out to about two disk scale radii is less than 0.2 per
cent of the total number of MBHs for all final halo masses. Part of the reason for this low number is that a lot
of the MBHs in the disk are orbiting at small distances of less than 1 per cent of the host virial radius and
are therefore counted as having fallen to the centre since their dynamics cannot be traced accurately any more
as mentioned above. Conversely the high mass end implies that apart from the central SMBH there will be one or
two other MBH of about a tenth of its mass orbiting in the halo

While the mass functions display a uniform slope of $N_{\bh} \propto M_{\bh}^{-1}$ it is interesting to note that
particularly in model A and C the mass function of MBHs in the most massive halo displays a downward departure
from a power law for MBH masses less than about $M_{\bh} \sim 3\times10^3 - 10^4$ \Msun. As a result the total
number of MBHs in the most massive halo is less than expected and is actually quite similar to that in the
second most massive halo in models A and C. This is due to the mass resolution limit of the semi-analytical code
as explained in section \ref{sec:semianalytical}. For the most massive final haloes this limit is larger and
consequently more haloes entering the hierarchy at this limit may already contain more than one seed MBHs, which
we then assume have merged. The latter means that the number of MBHs with the original seed mass is reduced.

\begin{figure*}
  \begin{minipage}{140mm}
    \centerline{\resizebox{\hsize}{!}{\includegraphics{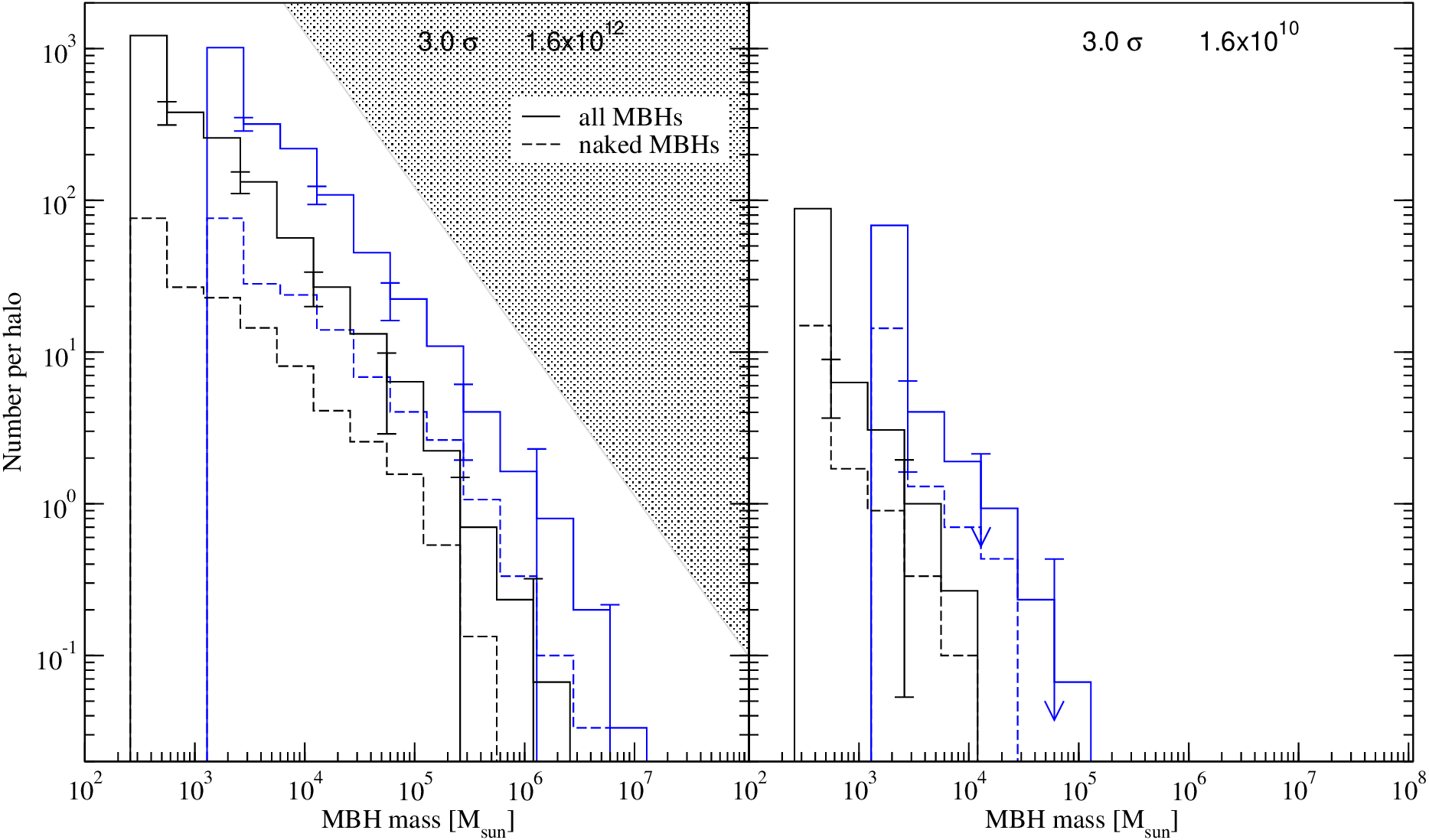}}}
    \caption{Comparison of models A and B: Abundance of all and naked
      MBHs in galactic haloes of mass
      $1.6\times10^{12}$ \Msun (left panel) and $1.6\times10^{10}$ \Msun
      (right). The set of curves to the left and right within each
      panel are
      for seed MBH masses of 260 and 1300 \Msun ~respectively. Provided the
      shape of the mass function remains the same for different MBH seed
      masses, the shaded area in the left panel indicates the prohibited regime where
      seed MBHs would need to be more massive than the total amount of
      baryons available in the original minihalo within which they formed.}
    \label{fig:MBHmassfunc_a_vs_b}
  \end{minipage}
\end{figure*}

In figure \ref{fig:MBHmassfunc_a_vs_b} we show the average MBH abundance including that of naked MBHs within the
virial radius of the $1.6\times10^{12}$ and $1.6\times10^{10}$ \Msun respectively for models A and B.

Compared to the mass of the bulge, disk and halo the seed MBH masses
are small and so we would not expect them
to significantly affect the evolution of substructure within the
host. For this reason we find that, except for
the high mass end, the MBH mass functions for the two different MBH
seed masses  are essentially the same but
are offset from one another along the ordinate (representing the
actual MBH mass) by a constant factor that is
more or less equal to the ratio of the initial seed MBH masses. Based
on this, the line bounding the grey shaded
area in figure \ref{fig:MBHmassfunc_a_vs_b} represents the inferred
mass function for a seed MBH with a mass of
$1.3\times10^{4}$ \Msun, that is the case where the entire baryonic
mass of an minihalo collapses  into the black
hole.

For a final halo mass of $1.6\times10^{12}$ \Msun ~in both model A and B we can deduce from figure
\ref{fig:MBHmassfunc_a_vs_b} that the number $N$ of remnant MBHs in the halo follows a power law
\begin{equation}\nonumber
N_{\bh} \propto M_{\bh}^{-1.01 \pm 0.04}
\end{equation}
which is also the basis on which we have determined the line bounding the prohibited region for any MBH mass
function. It is difficult to establish a similar uniform power law for haloes lighter than this. In this case not
as many massive MBHs have formed and the shape of the MBH mass function is thus dominated at the low mass end
(near the MBH seed mass) by the discrete nature of the MBH mass increase.

\begin{table*}
  \centering
  \begin{minipage}{100mm}
    \caption{Total mass in halo MBHs and mass of central SMBH
      for all final halo masses and all models. The total mass in MBHs is
      typically 2 -- 3 times larger than the mass of the central SMBH.}
    \label{tab:sum_MBH_Msmbh}
    \begin{tabular}{llll} \\ \hline \hline
      & &  \multicolumn{2}{c}{Halo mass} \\
      & & $1.6\times10^{10}$ & $1.6\times10^{11}$ \\ \hline
      model A & $\sum ~M_{MBH} [M_{\odot}]$ & $(3.4 \pm0.59)\times10^4$ &
      $(3.8 \pm0.62)\times10^5$ \\
      &$M_{SMBH} [M_{\odot}]$   & $(1.55\pm0.67)\times10^4$ &
      $(1.85\pm0.63)\times10^5$ \\
      model B & $\sum ~M_{MBH} [M_{\odot}]$  & $(1.34\pm0.36)\times10^5$ &
      $(1.68\pm0.32)\times10^6$ \\
      &$M_{SMBH} [M_{\odot}]$   & $(7.36\pm3.36)\times10^4$ &
      $(8.83\pm3.34)\times10^5$ \\
      model C & $\sum ~M_{MBH} [M_{\odot}]$  & $(6.48\pm0.98)\times10^4$ &
      $(4.79\pm1.49)\times10^5$ \\
      &$M_{SMBH} [M_{\odot}]$   & $(2.84\pm0.9 )\times10^4$ &
      $(3.62\pm1.5 )\times10^5$ \\
      model D & $\sum ~M_{MBH} [M_{\odot}]$  & $(5.75\pm1.84)\times10^3$ &
      $(5.63\pm0.99)\times10^4$ \\
      &$M_{SMBH} [M_{\odot}]$   & $(3.16\pm1.48)\times10^3$ &
      $(3.34\pm1.32)\times10^4$ \\
      \hline
      & & $1.6\times10^{12}$ & $1.6\times10^{13}$ \\ \hline
      model A & $\sum ~M_{MBH} [M_{\odot}]$ &   $(3.17\pm0.76)\times10^6$ &
      $(3.42\pm0.78)\times10^7$ \\
      &$M_{SMBH} [M_{\odot}]$   & $(1.5 \pm0.64)\times10^6$ &
      $(1.14\pm0.29)\times10^7$ \\
      model B & $\sum ~M_{MBH} [M_{\odot}]$  &   $(1.27\pm0.4 )\times10^7$ &
      $(1.78\pm0.31)\times10^8$ \\
      &$M_{SMBH} [M_{\odot}]$   & $(6.93\pm2.81)\times10^6$ &
      $(7.42\pm2.93)\times10^7$ \\
      model C & $\sum ~M_{MBH} [M_{\odot}]$  &   $(7.59\pm1.04)\times10^6$ &
      $(7.00\pm1.27)\times10^7$ \\
      &$M_{SMBH} [M_{\odot}]$   & $(3.23\pm1.21)\times10^6$ &
      $(2.96\pm1.29)\times10^7$ \\
      model D & $\sum ~M_{MBH} [M_{\odot}]$  & $(6.24\pm0.97)\times10^5$ &
      $(4.9 \pm1.08)\times10^6$ \\
      &$M_{SMBH} [M_{\odot}]$   & $(2.57\pm1.06)\times10^5$ &
      $(2.33\pm1.33)\times10^6$ \\  \hline\hline
    \end{tabular}
  \end{minipage}
\end{table*}
\begin{figure}
  \includegraphics[width=8cm]{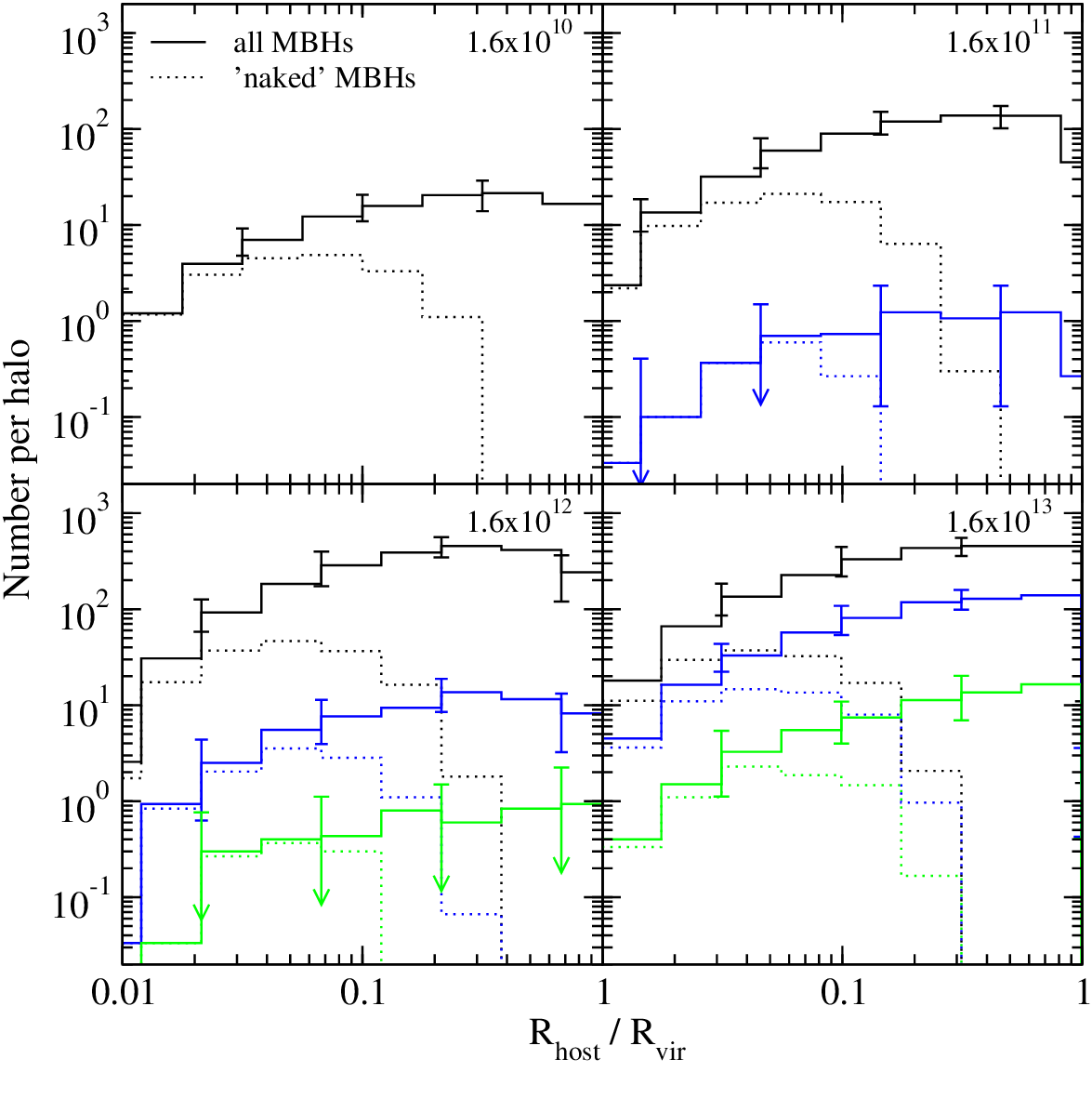}
  \caption{Radial distribution of MBHs for model A averaged over
    thirty trees for all four final
    halo masses. The contribution to the total of naked MBHs is
    shown by the dotted lines. In each panel the two lower sets of
    curves represent
    the radial distribution for MBHs more massive than $10^4$
    and $10^5$ \Msun respectively. The relative distributions of
    MBHs above the different mass thresholds are similar, implying that there
    is no obvious mass segregation.
  }
  \label{fig:abund_radius}
\end{figure}
Figure \ref{fig:abund_radius} shows the number of MBHs as a function of distance from the host centre for model
A only. We have also plotted the contribution of MBHs more massive
than $10^4$ and $10^5$ \Msun. In addition the plot shows the fraction of naked MBHs for all of the above cases. As
expected there are more naked MBHs near the centre where the steeper host gravitational potential results in
stronger tidal forces that strip away more matter from any satellite present there. In all cases shown the
contribution of naked MBHs typically becomes dominant at distances of less than a few percent of the host virial
radius. With increasing halo mass the relative contribution of naked MBHs at small radii decreases. A reason for
this is that in small haloes the tidal gradient is relatively steeper and MBHs are therefore more likely to be
stripped. For the different MBH mass cuts, however, there does not appear to be any significant difference in
the relative radial distribution of naked MBHs.

\begin{figure}
  \centering
  \centerline{\resizebox{\hsize}{!}{\includegraphics{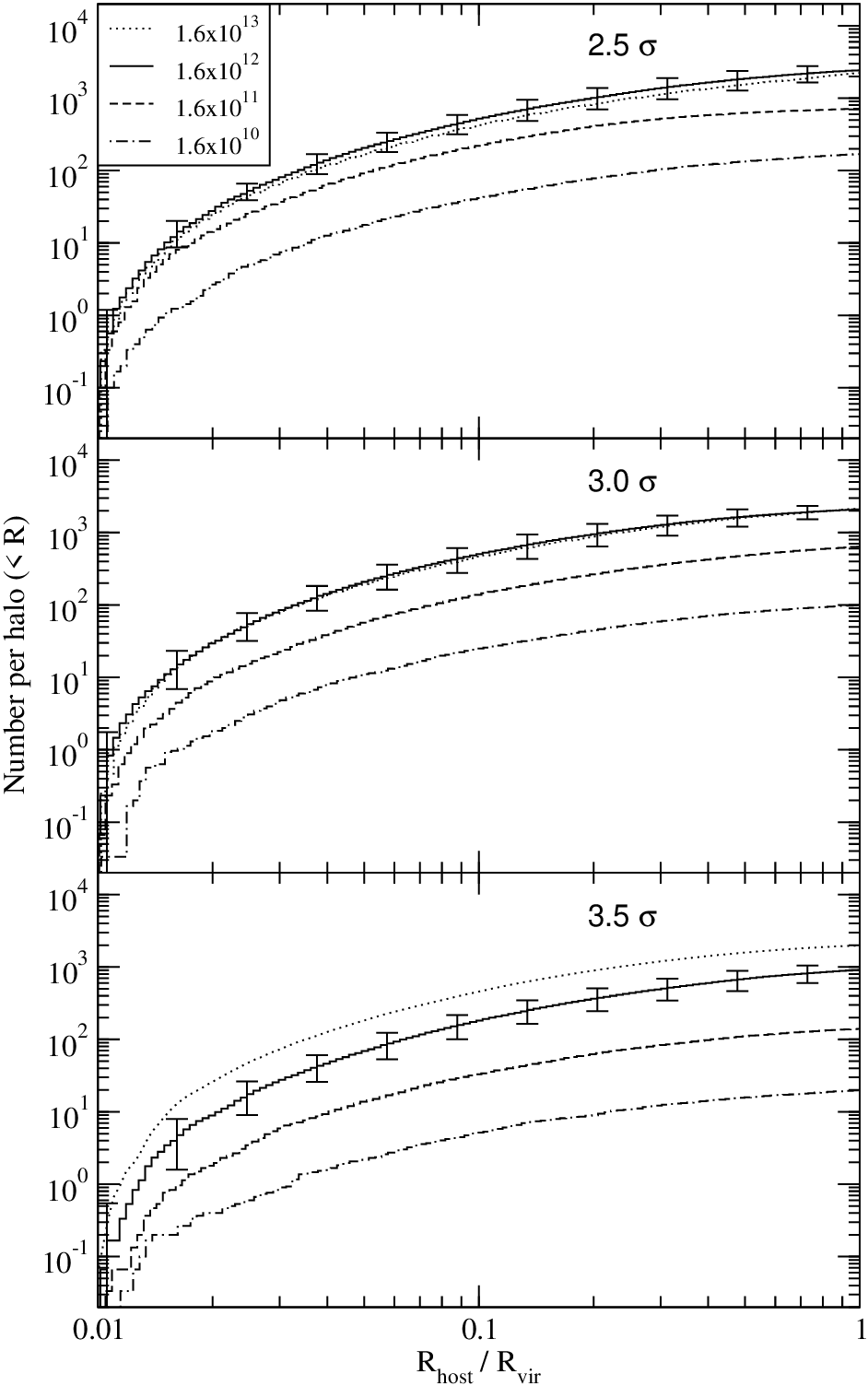}}}
  \caption{Number of MBHs below radius R from host centre, averaged
      over all trees and scaled to the respective virial
      radius for the individual haloes. Error bars corresponding
      to the standard deviation are shown for the case of the $1.6
      \times10^{12}$ \Msun ~halo.}
  \label{fig:cumul_abund_radius}
\end{figure}

In figure \ref{fig:cumul_abund_radius} we have shown the cumulative abundance of MBHs below a given distance
from the halo centre. The distribution is scaled to the virial radii of the respective haloes and results are
shown for models A,C \& D. What this shows us is that the shape of the radial distribution is fundamentally the
same in all cases except for the normalisation. The latter corresponds to the difference in the total MBH
abundance as shown in figure \ref{fig:MBHmassfunc}. We also note that, while different in all other cases, the
radial distribution of MBHs are very similar again for final halo masses of $1.6\times10^{12}$ and
$1.6\times10^{13}$ \Msun in models A and C. This is for the same reasons given above in respect of the
corresponding mass functions.

An important result that emerges from both figures \ref{fig:MBHmassfunc} and \ref{fig:cumul_abund_radius} is
that both the shape of the mass abundance as well as the radial distribution of MBHs in haloes is very similar.
The different normalisation of the number of and mass in MBHs for the different halo masses and models is
summarised in tables \ref{tab:global_abund} and \ref{tab:sum_MBH_Msmbh}. For comparison the latter also lists
the mass of the respective central SMBHs. In section \ref{sec:SMBH_abund} we will see that the mass of SMBHs is
tightly correlated with the mass of the galactic bulge component. The table shows that the mass contained in
halo MBHs is typically between two to three times larger than the mass of the central SMBH.

Together with the correlations between $M_{SMBHS}$ and $M_{bulge}$ and between $M_{bulge}$
and $M_{halo}$ this implies that a similar correlation exists between the total mass contained in MBHs and their
respective halo masses. In fact, for all four halo masses this correlation is consistent with
\begin{equation} \nonumber
  \sum ~M_{\bh} \propto M_{halo}^{1.0\pm0.03}
\end{equation}
which is what we would expect, since any halo increasing its mass through the accretion of smaller haloes will
also inherit all MBHs associated with the latter.

Within the standard deviation quoted we expect the number and mass abundance of MBHs particularly in the
$1.6\times10^{12}$ \Msun halo to be representative of Milky-Way-sized galaxies in currently favoured
$\Lambda$CDM cosmologies.

In table \ref{tab:local_abund} we have listed the average abundance of MBHs in a local Earth-centred volume
(which we have taken as corresponding to a volume centred at 8.5 kpc from the centre of a $1.6\times10^{12}$
\Msun ~halo in our simulations). Virtually all of these will be seed BHs that have not yet merged and in the
absence of any growth process other than hierarchical merging their mass will be equal to that of the initial
seed BHs.
\begin{table}
\centering \caption{Abundance of MBHs in Earth-centred volumes at
8.5 kpc from the galactic centre in the Milky-Way-sized halo ($1.6\times10^{12}$ \Msun). Given are the average
over thirty trees with their respective standard deviation.} \label{tab:local_abund}
\begin{tabular}{llll}  \\ \hline\hline
   & \multicolumn{3}{c}{Distance from Earth $\Delta r$ [kpc]}\\
 & 2.0  & 2.5  & 3.0  \\ \hline
Model A & 0.87 $\pm$ 0.94 & 1.7  $\pm$ 1.29 & 2.57 $\pm$ 1.76\\
Model B & 0.43 $\pm$ 0.63 & 1.03 $\pm$ 0.96 & 2.00 $\pm$ 1.56\\
Model C & 0.53 $\pm$ 0.73 & 1.4  $\pm$ 1.28 & 2.47 $\pm$ 1.87\\
Model D & 0.23 $\pm$ 0.63 & 0.4  $\pm$ 0.72 & 0.70 $\pm$ 0.84\\ \hline\hline
\end{tabular}
\end{table}
To the extent that the large standard deviations allow for any meaningful comparison between the models, one
notable feature is the marked difference between the average number of nearby MBHs in models A and B. We already
mentioned that we would not expect any significant difference in the halo (merger) dynamics and thus the final
MBH abundance because the models only differ in that they start out with different MBH seed masses. However, the
solar neighbourhood at 8.5 kpc from the centre is well within the radial regime where the MBH abundance is
becoming increasingly dominated by naked MBHs. At distances smaller (larger) than the sun's orbit there are
relatively more (less) naked MBHs and because these are more massive they move towards the centre more quickly.
This would explain why the number of nearby MBHs is consistently lower in model B for all local distances
considered.
Our findings for the MBH abundance in haloes are in accord with the results of another recent investigation by
Volonteri, Haardt \& Madau (2003, hereafter VHM03) \nocite{volonteri03}. We find that the total MBH mass density
in a Milky-Way sized galactic in our model C (with a minihalo collapse threshold of 3.5 $\sigma$) halo agrees to
within a factor 2 with their value inferred from the density function of `wandering' BHs in galactic haloes.
\subsubsection{Constraints on initial MBH mass function}
Figure \ref{fig:MBHmassfunc_a_vs_b} for the two
different MBH seed masses gives some indication of the effect of other changes in the masses and numbers of seed
MBHs in the primordial haloes.

We have seen above that the MBH mass functions are shifted along the $M_{MBH}$ axis in proportion to the mass of
the seed MBHs. This mass, however, cannot be higher than the total
baryonic mass contained in the original minihaloes. This translates
into the grey shaded area shown in figure \ref{fig:MBHmassfunc_a_vs_b} and thus any mass
for a single seed MBH between 260 and $1.3\times10^4$ \Msun will lead
to a present-day MBH mass function that
lies between the shaded area and the mass function corresponding to model A.

By conservation of mass \footnote{Strictly the masses of two merging BHs are not conserved, but will be lower by
a few per cent, since gravitational waves can radiate away some of the BHs' rest mass energy. In the following
we assume that this effect only changes our results by a negligible amount, although the mass loss through
gravitational radiation accumulated in many mergers for some MBHs may become significant.}, if the primordial
halo contains more than one MBH of different masses in the range 260 \Msun $< M_{\bh} <$ 13000 \Msun then the
resulting mass function will again lie between the bottom and the top one shown, but will have a different
slope. If initially one or more MBHs were present with masses lower than $260 {\rm M_{\odot}}$, the present-day
mass function will correspondingly extend to lower masses, but will otherwise still be limited by the shaded
area. This means, that even though we had initially made a fairly specific choice for the initial MBH mass
function in the primordial haloes, any general form for the MBH IMF is expected to lead to results within the
limits provided by the MBH mass functions shown, if there is at least one seed MBH of $260$ \Msun or larger.
This is {\em provided the
  seed MBHs form in $3\sigma$ peaks} as is the case for model A and B. Seed MBH formation in
higher or lower peaks changes the overall normalisation and leads to a corresponding scaling of the mass
functions. The relative range of possible mass functions should nevertheless remain reasonably well defined
unless seed MBHs form in minihaloes collapsing from an extended range of peak heights and thus redshifts.

We need to stress that the above depends on the  assumption that all MBHs falling to within one per cent of the
virial radius merge efficiently in all haloes merging along the way to produce the final host halo.

If the only or at least most significant source of seed MBHs is that forming in minihaloes then the total mass
contained in halo MBHs can be used to normalise the initial mass function of seed MBHs, to which it is related
by the background cosmology. The latter determines the average merger history of haloes and thus the average
number of minihaloes ending up in more massive haloes later on. Note that this is not much affected by the merger
efficiency of MBHs since the present-day MBH mass function is dominated by seed MBHs that have not merged, and
that contribute a similar amount to the total mass contained in halo MBHs as the few very massive MBHs that have
resulted from multiple mergers of seed MBHs. This is just expressing in a different way the $N_{\bh} \propto
M_{\bh}^{-1}$ scaling we found earlier (c.f. figure
\ref{fig:MBHmassfunc_a_vs_b}, which means that the total mass contributed from successive logarithmic
MBH mass intervals is constant.

\subsection{MBH mass accretion rates} \label{sec:MBH_gas_accretion}
Having established the abundance of MBHs we now look at rates at which
MBHs accrete material from their surroundings. This
forms the basis for any estimate of the accretion luminosities that
may be detectable.
\subsubsection{Bondi-Hoyle accretion}
If we consider a MBH travelling within a uniform medium, the steady state
accretion rate is given by the Bondi-Hoyle accretion rate \cite{bondi44,bondi52}
\begin{equation}
    \frac{\d M}{\d t} = \pi r_{acc}^2 \sqrt{v_{\bullet}^2 + c_s^2} \rho_g
\end{equation}
Here $c_s$ is the sound speed in the gas and $\rho_g$ its density -- both far from the MBH. $v_{\bullet}$ is the
velocity of the MBH and $r_{acc}$ the accretion radius
\begin{equation}
    r_{acc} = \frac{2 G M_{\bullet}}{v_{\bullet}^2 + c_s^2}
\end{equation}
giving an accretion rate
\begin{equation}
    \frac{\d M}{\d t} = \frac{4 \pi G^2 M_{\bullet}^2 \rho_g}{c_s^3} ( 1 + \beta_s^2)^{-3/2}
\end{equation}
where we have used $\beta_s \equiv v_{\bullet}/c_s$ \cite{chisholm02}. If the MBH accretes adiabatically from a
gas of pure hydrogen this is
\begin{eqnarray}
    \dot{M} &=& 8.77 \times 10^{-12} \left(\frac{M_{\bullet}}{100
M_{\odot}}\right)^2 \nonumber \left(\frac{\rho_g}{10^{-24} {\rm g
~cm}^{-3}}\right) \\
& & \times \left(\frac{c_s}{10 {\rm ~km ~s}^{-1}}\right)^{-3} (1 + \beta_s^2)^{-3/2} ~M_{\odot} {\rm yr}^{-1}
\end{eqnarray}
It is implicit that only baryonic matter can get accreted in this way, assuming that a mechanism exists for the
dissipation of thermal energy as the material falls towards the MBH
Any such process is not relevant for dark matter, which is by
definition not or only weakly interacting and only gets accreted if it approaches the MBH to within a
distance that is of the order of the last stable orbit of the MBH. This is much smaller - i.e. it has a much
smaller `cross section' - than the Bondi accretion radius and we will subsequently neglect the possibility of
dark matter accretion.

In what follows we assume that while the nature of the accretion process is somewhat uncertain, the overall mass
accretion rate is essentially determined by the Bondi-Hoyle formula. This implies, that we neglect the
possibility that the mass accretion rate is modified e.g. by a non-negligible mass of an accretion disk that may
form around the accreting MBHs.
\subsubsection{Accretion environment: ISM vs. baryonic core
  remnants}
\label{sec:baryoniccoreaccretion}
As the MBHs orbit through the host halo they accrete matter from the host ISM. This will only be significant in
regions with relatively large amounts of gas, which is the case primarily in the galactic disk and bulge.

Alternatively the MBHs may accrete from a core remnant of the
satellite they were originally associated with. In
our numerical procedure a satellite is considered tidally disrupted
when the tidal radius becomes smaller than
the scale radius of the satellite density profile. We have so far
referred to MBHs embedded in such satellites as `naked' MBHs. This condition is
sufficient as far as the dynamical
importance of the satellite is concerned, since at this stage it will
have lost all but at most a few percent of
its mass. However, even in these systems a core remnant close to the satellite centre may still
remain. While insignificant for the
overall dynamics of the satellite the amount of baryonic matter
contained in this core could still contribute
significantly to the accretion onto a MBH that is present at its centre and so
boost its accretion luminosity
potentially by orders of magnitude. In this case MBH accretion is
independent of the conditions and relative MBH
velocity in the surrounding ISM of the host halo.

To determine the mass accretion rate we still need an estimate of the gas density in the baryonic core that is
essentially acting as fuel supply travelling along with the MBH. Assuming that all baryonic matter has cooled in
the satellites before they are subject to tidal stripping and heating in the host the outer radius of the
baryonic core assumed spherical is $r_{b} \sim 0.1 r_{vir}$ where $r_{vir}$ is the virial radius of the
unstripped satellite halo of mass $M_{vir}$. If all baryons are in the form of gas and the baryon fraction in
the satellite is cosmological, i.e. $M_{b} = (\Omega_b/\Omega_m) ~M_{vir}$, then the mean gas density is
\begin{equation}
\rho_g =  \frac{3 M_{b}}{4 \pi r_{b}^3} =\frac{1}{0.001} ~\frac{\Omega_b}{\Omega_m} ~\frac{3 M_{vir}}{4 \pi
r_{vir}^3} \approx
 2.39 \times 10^{2} \frac{\Omega_b}{\Omega_m}~\frac{M_{vir}}{r_{vir}^3}
\end{equation}
Substituting this into equation (\ref{eq:Lbol}) and using $c_s \approx 10 {\rm ~km ~s}^{-1}$ for ISM at $10^4$
Kelvin we can determine the mass accretion rate
\begin{eqnarray}
    \dot{m} & \approx & 6.25 \times 10^{-8} \left(\frac{M_{MBH}}{M_{\odot}}\right)
\frac{\Omega_b}{\Omega_m} \left(\frac{M_{vir}}{10^5
M_{\odot}}\right) \nonumber \\
& & \times \left(\frac{r_{vir}}{{\rm kpc}}\right)^{-3} \left(\frac{c_s}{10 {\rm ~km ~s^{-1}}}\right)^{-3}
\end{eqnarray}
and thus the luminosity. Here $\dot{m}$ is the mass accretion rate in units of the Eddington accretion rate
\begin{equation} \label{eq:dimless_accrate}
  \dot{m} \equiv \frac{\dot{M} c^2}{L_E} = 1.53
  \left(\frac{\dot{M}}{10^{17} {\rm g ~s}^{-1}}\right) \left(\frac{M_{\bh}}{M_{\odot}}\right)^{-1}
\end{equation}

\subsubsection{Distribution of mass accretion rates}
\begin{figure}{}
  \centering
  \centerline{\resizebox{\hsize}{!}{\includegraphics{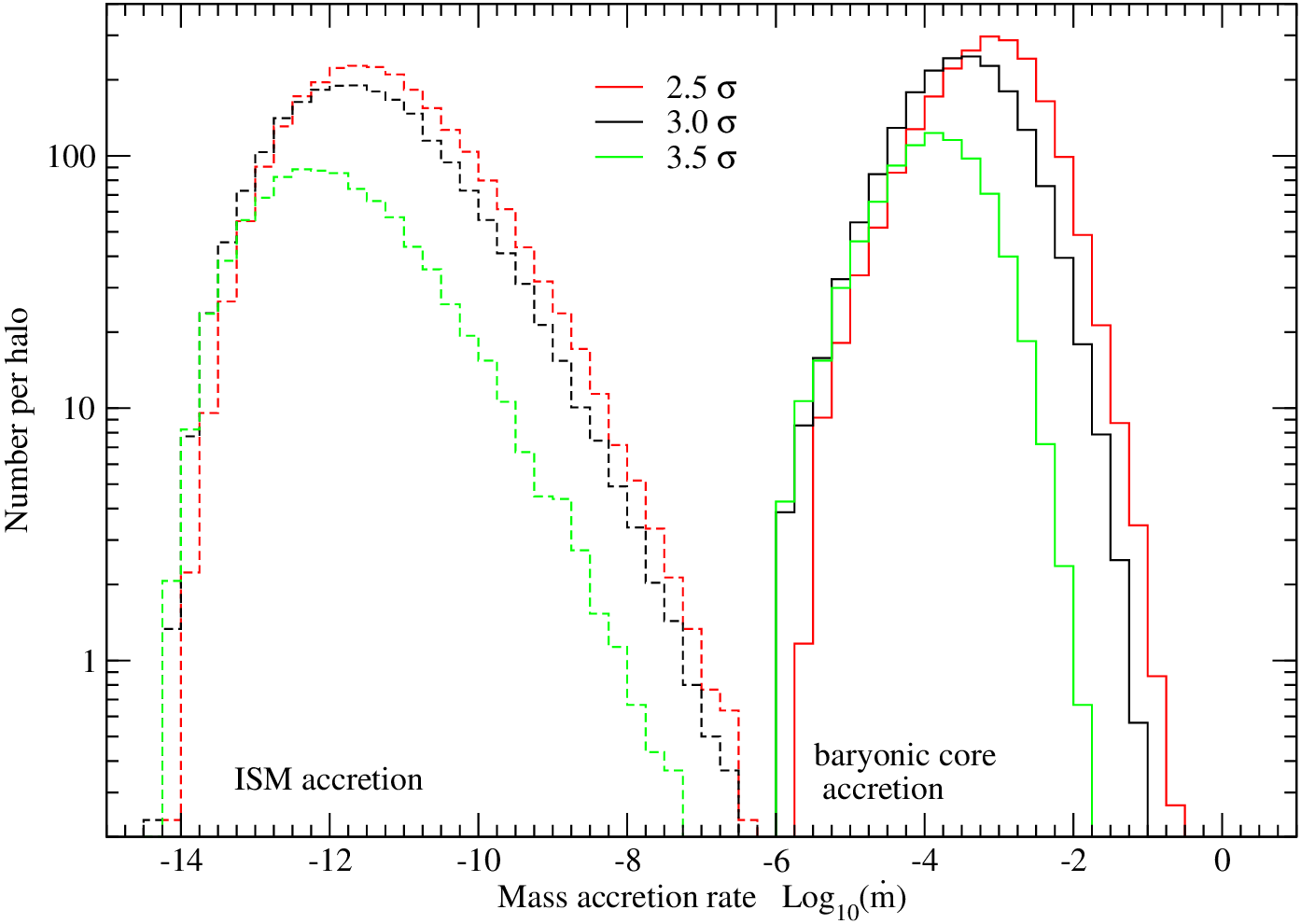}}}
  \caption{Abundance of accretion rates for MBHs in a Milky-Way sized
    halo for models A, C and D. The curves on the left correspond to ISM
    accretion, the ones on the right to baryonic core accretion. }
  \label{fig:accrate}
\end{figure}

In figure \ref{fig:accrate} we have plotted the number of MBHs vs their respective mass accretion rates for ISM
as well as baryonic core accretion. The plot serves primarily to highlight the vast difference in accretion
rates for the two models. In general the accretion rates for ISM accretion are lower than that for baryonic core
accretion by some 7 - 10 orders of magnitude!
We can see that for the case of ISM accretion the maximum accretion rate does not exceed $\sim 10^{-7}$ of the
Eddington value. It seems then that ISM accretion in the context of our model is completely insignificant. There
are two reasons for this. Firstly, MBHs are distributed across the halo as we have seen above, with the number
of MBHs in the disk and bulge small and certainly very much lower than that of stellar mass BHs. Secondly, even
in the disk and bulge the actual density of accretable gas is  too low especially at late times. In our analysis
of the observable signatures of accreting MBHs we therefore mainly focus on baryonic core accretion for which we
get consistently larger accretion rates.

\begin{figure}
  \centering
  \centerline{\resizebox{\hsize}{!}{\includegraphics{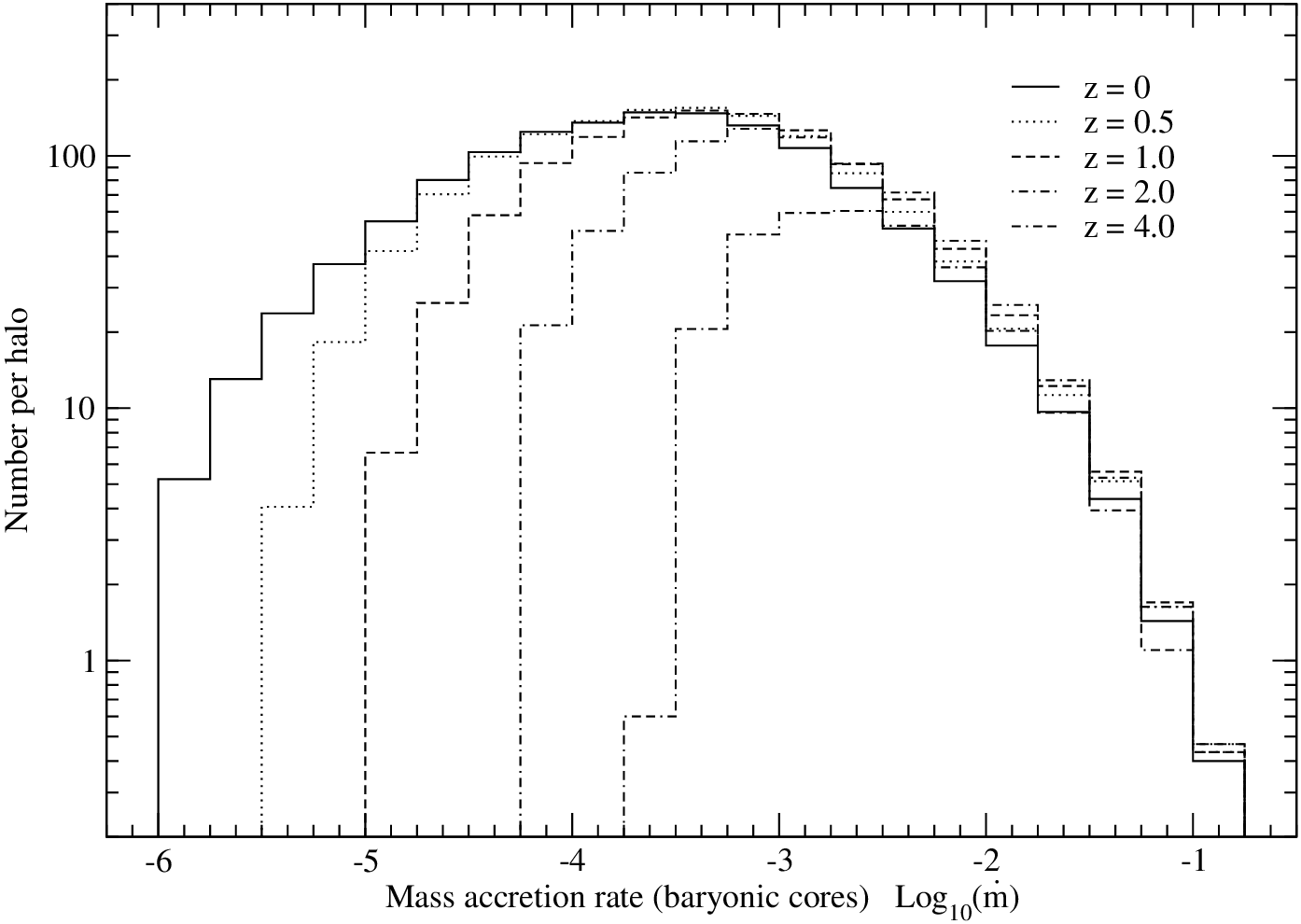}}}
  \caption{Redshift dependence of baryonic core accretion in model A
    and a final halo mass of $1.6\times10^{12}$\Msun. At late
    times the host virial radius becomes largest. The
    resulting inclusion of low mass haloes with correspondingly
    small MBHs leads to an increase of objects with low
    accretion rates. (This particular plot is based on a
    different set of simulations which does, however, use the same
    key cosmological parameters.)}
  \label{fig:z_accrate}
\end{figure}
For model A we show in figure \ref{fig:z_accrate} the average distribution of mass accretion rates of MBHs for
various redshifts in a halo that grows to mass $1.6 \times 10^{12}$ \Msun at $z = 0$. We see that for increasing
redshift the lower end of the range of accretion rates moves towards higher values. The reason for this is that
at late times, as the host virial radius becomes larger, more satellites are being incorporated in the outer
parts of the host. Most of these will be small satellites with seed mass MBHs

It is also interesting to note that the maximum accretion rate in all cases does not exceed a value of more than
about 10 \% of the Eddington
 mass accretion rate.
Assuming that the maximum accretion rates have never been larger than this it is obvious that gas accretion from
baryonic cores for most MBHs does not lead to significant mass increase over a Hubble time.
\begin{figure*}
  \begin{minipage}{160mm}
    \centering
    \includegraphics[width=12cm,height=15cm]{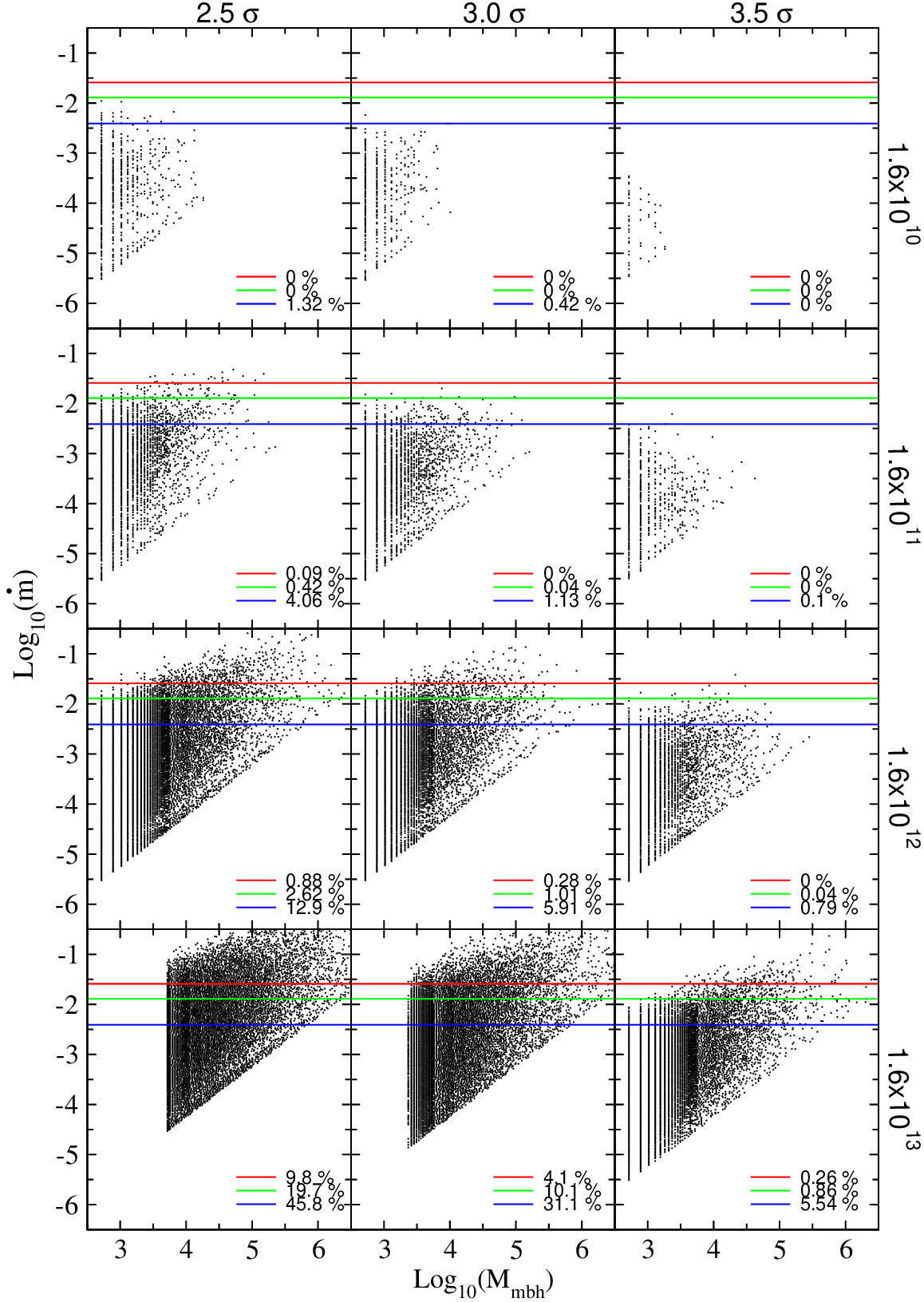}
  \end{minipage}
  \caption{Dimensionless baryonic core accretion rates in Eddington units vs. MBH
    mass for all final halo masses in models C, A and D (from
    left to right). The results shown are for all 30 trees. In each
    panel the horizontal lines from top to bottom denote the accretion
    rates above which the MBH mass would grow by at least a
    factor 100, 10 and 2 respectively, within a Hubble time, provided the
    accretion rate stays constant. The percent figures show the
    fraction of MBHs with accretion rates higher than the
    respective cut-offs.}
  \label{fig:eddington_accrate}
\end{figure*}
This is shown in figure \ref{fig:eddington_accrate} where we have plotted the accretion rates in Eddington units
for all final halo masses in models A, C and D. In particular we have marked various fractions of MBHs that
accrete at rates such that their mass would increase by a factor 2, 10 and 100 respectively within a Hubble time
provided the rate stays constant. Except for the most massive haloes not more than about 10 percent of MBHs
accretes even enough to double their mass, while in all cases no more than about 20 percent grow in mass by more
than a factor of 100. The factor 2 mass increase is of special importance: In the most conservative baryonic
core accretion scenario we would expect an MBH to be embedded in a core that contains an amount of material at
least of the order of the initial MBH mass. This is to say the MBH holds on to the material within its initial
range of influence. If this applied to all MBHs we would have to discount any MBHs accreting at a rate higher
than needed to double their mass within a Hubble time - call this $\dot{m}_{2\times}$, since they would have
consumed their core by now. We will not undertake a selection of presently accreting MBHs on the basis of this
criterion, because accretion rates for most MBHs are clearly below $\dot{m}_{2\times}$. We also assume that a
baryonic core is significantly more massive than the mass of the MBH. This is plausible as the baryonic
component has condensed at the centre of haloes and thus is much less affected by any tidal stripping of the
outer parts of the halo which are dominated by dark matter.

In the context of our baryonic core accretion model realistic accretion rates are likely to be significantly
lower, especially at low redshift, when a lot of gas has already been used up in star formation - an effect that
we have not accounted for in our simulations.

The distribution of data points for the most massive haloes in figure \ref{fig:eddington_accrate} can be fit by a
power law
\begin{equation}\label{eq:accrate_vs_MBHmass}
  \dot{m} \propto M_{\bh}^{0.68 \pm 0.02} \Rightarrow \dot{M} \propto M_{\bh}^{1.68 \pm 0.02}
\end{equation}
i.e. the most massive MBHs are also those accreting at the highest rates in general. Due to the scatter in the
plots some of the largest dimensionless accretion rates do actually occur for MBHs that are one or two orders of
magnitude lighter than the most massive ones. However, the largest physical accretion rates \footnote{These
  are not scaled to the MBH mass as is the case for the dimensionless
  accretion rates.} and thus accretion luminosities are indeed those
of the most massive MBHs.

\subsection{Abundance and mass of SMBHs} \label{sec:SMBH_abund}
MBHs that move at distances from the centre less than 1 \% of the host virial radius are considered as having
fallen to the centre (c.f. section \ref{sec:mergeff}). In line with the assumption of efficient
merging the mass of these infalling MBHs is simply added to the mass of a single SMBH growing at the centre.
\subsubsection{SMBH from hierarchical merging of remnant MBHs}
\begin{figure*}{}
  \begin{minipage}{140mm}
      \centerline{\resizebox{\hsize}{!}{\includegraphics{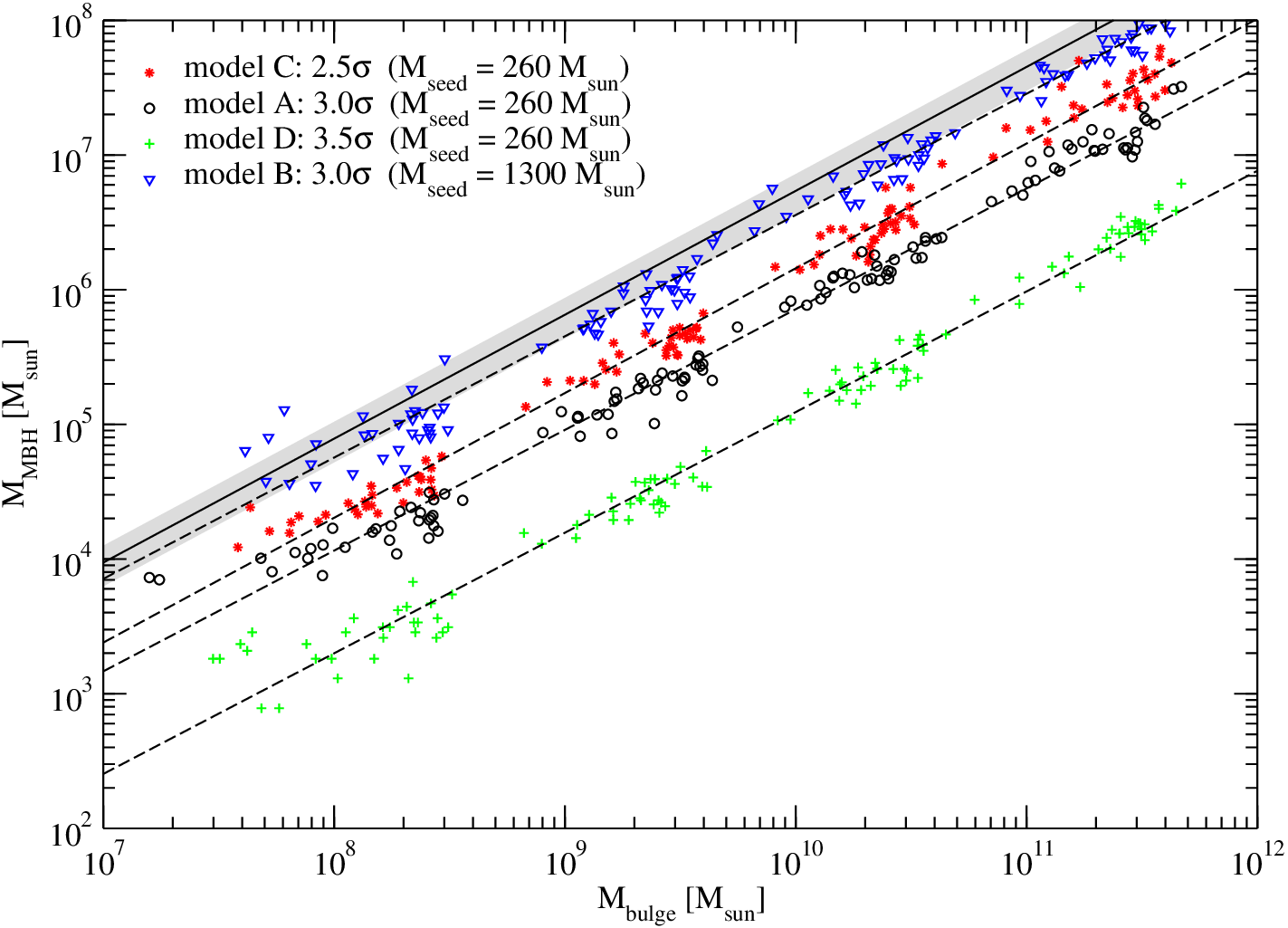}}}
  \end{minipage}
  \caption{Central SMBH
    masses vs. bulge mass for all seed MBH
    masses and peak heights considered (models A -- D). For each combination the
    data for all corresponding final halo masses have been grouped
    together. Dashed lines represent the best fits to the data. The
    solid line and shaded area are the $M_{SMBH} - M_{bulge}$
    relation and associated error from Magorrian ~\ea 1998.}
  \label{fig:MsmbhMbulge}
\end{figure*}

Figure \ref{fig:MsmbhMbulge} shows the relation
between the mass of the galactic bulge and the central SMBH if the latter grows purely through mergers of
smaller MBH. The solid line represents the linear relationship between SMBH and bulge mass as determined from
observations. To determine this we have used the MBH mass - bulge luminosity relation based on more recent
compilation of data \cite{kormendy01}
\begin{equation}
  \left(\frac{M_{\BH}}{M_{\odot}}\right) = 1.24 \times 10^{-3}
  \left(\frac{L}{L_{B,\odot}}\right)^{1.08}
\end{equation}
where $M_{\BH}$ denotes the mass of the SMBH. We have then combined this with the mass to light ratio determined
by Magorrian ~\ea (1998) \nocite{magorrian98}
\begin{equation}
  \left(\frac{M_{\BH}}{M_{\odot}}\right) = 0.33\pm0.11
  \left(\frac{L}{L_{\odot}}\right)^{1.18 \pm 0.03}
\end{equation}
We also assumed that the $B$ band luminosity is approximately the same as the bolometric luminosity $L_{B,\odot}
\approx L_{\odot}$.
The observed relation places an upper limit on the allowed masses of seed MBHs and peak heights in the initial
density field. The normalisation of the $M_{\BH} - M_{bulge}$ mass relation in figure \ref{fig:MsmbhMbulge} is
primarily a function of the mass density of MBHs which, in the absence of accretion, is defined by the number
and mass of the seed MBHs in our model. For model B no significant gas accretion is required to match the
observed relation and even for models A and C accretion would only be required to increase the SMBH mass by less
than a factor of 10. Only in model D (also the fiducial model of VHM03) gas would actually have to raise the
SMBH mass by about a factor 100.

The slope of the observed relation, $M_{\BH} \propto M_{bulge}^{0.92 \pm 0.02}$ is very similar to those of the
best fits to our data. All the best fits in figure \ref{fig:MsmbhMbulge} have $M_{\BH} \propto M_{bulge}^{0.90
\pm
  0.01}$, except for the model with 2.5$\sigma$, $M_{\bh,seed} = 260
~M_{\odot}$, for which we get $M_{\BH} \propto M_{bulge}^{0.93 \pm
  0.01}$. This agreement is surprising, especially since one would
intuitively expect a simple linear relation between $M_{\BH}$ and $M_{bulge}$ in our merger only scenario. On
the other hand, this close agreement raises the question of whether the observed slope is primarily a relic of
the hierarchical merger process rather than the result of accretion processes.

To the extent that gas accretion occurs it does not alter the slope, i.e. all SMBHs grow through gas accretion
by the same factor. At first sight this would seem to be in contradiction to the result for halo MBHs that more
massive MBHs systematically accrete at higher rates (c.f. figure \ref{fig:eddington_accrate}). We can resolve
this in the same way as for halo MBHs: At low redshifts a significant if not the largest amount of accretable
gas may have already been consumed in star formation.

For the SMBHs our results are also consistent with that of VHM03. For model A, for instance, we found the mass
of a central SMBH in a Milky-Way sized halo to be $1.5 \times 10^6$ \Msun. Accounting for the difference in seed
MBH masses used this agrees with the central SMBH mass of $\sim 1\times 10^6$ \Msun for a halo of mass
$1.6\times10^{12}$\Msun with $\sigma \sim 155 {\rm ~kms}^{-1}$ as implied by their $M_{\BH} -\sigma$ relation
(with no gas accretion). This also coincides with the mass determined for the SMBH in the Milky Way, although
the Milky Way SMBH is known to lie significantly below the observed $M_{\BH} - \sigma $ relation.

However, our slightly non-linear $M_{\BH} - M_{bulge}$ correlation corresponds to a $M_{\BH} - \sigma $ relation
whose logarithmic slope ($\sim 4.0$) does not match the much flatter one they determined ($\sim 2.9$) for $3
\sigma$ collapse and no gas accretion. We believe this to be possibly a result of the different assumptions made
about the MBH merger process. In particular the inclusion of triple BH
interactions and sling-shot ejections by VHM03 would probably lead to even lower central SMBH masses in our analysis.

\subsubsection{SMBH growth from gas accretion}
A number of studies \cite{soltan82,yu02} suggest that the present day SMBH mass density is consistent with the
amount of gas accreted during the optically bright QSO phase. If this is the case then our model B is probably
ruled out as it only requires a mass increase of order unity to match the observed abundance of SMBHs today.
Models A, C and D would not be affected as gas accretion would still be needed to increase the SMBH masses by at
least a factor 3.

On the other hand gas accretion (during the QSO phase) alone cannot explain growth from stellar mass BHs to
the most massive SMBHs ($\ge 10^9$ \Msun). Even if stellar mass BHs are accreting at the Eddington limit, the
QSO phase would not last long enough for the BHs to grow sufficiently.

If BHs accrete at the Eddington limit their mass doubling time is
given by the Salpeter time scale,
\begin{equation}\label{eq:salpetertime}
t_{salp} \sim \frac{\eta ~M ~c^2}{L_{edd}} = \frac{\eta ~\sigma_t c}{4 \pi
G m_p} \sim \eta ~4.5 \times 10^8 {\rm yr}
\end{equation}

The quasar epoch lasts about $10^9$ years from a redshift of about $z \sim 3.5
- 1.5$ \cite{richstone98}. Typical QSO lifetimes are estimated to be $t_Q \sim 10^{7} - 10^8$ yrs. Even if we
assume that a series of major mergers has triggered individual QSOs repeatedly such that they are active through
most of the QSO epoch, the BHs would only grow by a factor of order $10^6$ but likely less.

Semi-analytical models of galaxy formation assume that following the QSO epoch the (S)MBHs powering quasars grew
further through major mergers \cite{haehnelt00}. However, the latter is unlikely to have raised the (S)MBH mass
by more than an order of magnitude. This means that MBHs with masses of at least $10^2 - 10^3$ \Msun must have
been present at the beginning of the QSO epoch already. This is confirmed by the presence of luminous quasars at
redshifts higher than $z \sim 6$. In fact, the latter means that at least some SMBHs must have been in place by
that time as the Universe was hardly old enough to accommodate a long enough Eddington limited gas accretion
phase to explain their mass \cite{haiman01}.

Our model complements this idea. Hierarchical merging involving seed MBHs originating at $z \grtsim 20$ can
produce the number and masses of (S)MBHs needed at the onset of the QSO epoch. To test this idea we estimate the
density of SMBHs with a mass of at least $10^8$ \Msun  - which we take as the mass required to power a QSO - at
a redshift $z \sim 6$. Since only some fraction of QSOs are active, the number density of SMBHs has to be at
least as large as the density of luminous QSOs at that redshift, which is $\sim 10^{-7} {\rm Mpc}^{-3}$
\cite{richstone98}. This number density corresponds to the abundance of haloes of mass $6\times10^{12}$ \Msun at
redshift $z = 6$ in the $\Lambda$CDM model we are working with. If we assume that the $M_{\BH} - M_{bulge}$
relation in figure \ref{fig:MsmbhMbulge} and the $M_{\BH} - M_{host}$ relation in table \ref{tab:sum_MBH_Msmbh}
will be the same at $z = 6$ then there is a problem: For a halo mass of $6\times10^{12}$ \Msun models A and C
imply a SMBH mass of $5\times 10^{6}$ to $1\times 10^{7}$ \Msun. However, if we allow for gas accretion at only
a fifth of the Eddington rate these SMBHs could have certainly grown by an order of magnitude by $z = 6$. We
conclude then that Eddington limited gas accretion alone onto an initially stellar mass BH or even a
moderate mass MBH cannot produce the number density of SMBHs required to explain QSOs at redshifts
of $z \sim 6$.
The need for intermediate mass seed BHs and/or some merging of MBHs/SMBHs is therefore necessary to explain the
presence of the most massive SMBHs.

\subsection{Effect of low merger efficiency}
If the merging of MBHs does not proceed efficiently, our results
presented above will be affected in two major respects.
In our model the build-up of SMBHs requires the merging of MBHs at the
centre of merging haloes. If merging does not occur efficiently, SMBHs
would have to grow by other processes, such as gas accretion.
Similarly MBHs with masses larger than the MBH seed mass would be
much harder to form, i.e. the number of the most massive MBHs orbiting
in haloes would be very much lower.
Since the MBH accretion rates depend also on halo mass, this means
that the number of MBHs with high accretion rates and accretion
luminosities will be reduced, too.

Interestingly -- and contradictory at first sight -- less efficient
merging opens up more possibilities for
processes to aid the merging of MBHs.
If merging is less efficient, gas accretion will have to play an even
more prominent role in MBH growth.
It could be argued that, if gas accretion is a generic and decisive
constituent of MBH growth even at high redshifts, the resulting mass increase of MBHs would
enhance the subsequent merging of MBH binaries.
Secondly, less efficient merging implies that three body interactions
can occur. Typically, if three or more MBHs interact, the lightest
ones will be slingshot-ejected from the system, leaving behind a binary that is yet
more tightly bound and will merge in a shorter time. This is the
same process as for stellar dynamical
interactions mentioned in section \ref{sec:mergeff}, albeit with a
much higher efficiency per interaction.

In one important respect, however, a lower merger efficiency has no
significant effect. We have seen above that the number of MBHs in haloes is
inverse proportional to the MBH mass. The number of halo MBHs is
therefore dominated by seed MBHs, and a change in merger efficiency
would not significantly affect the total number of MBHs in haloes.

\section{Emission signatures from MBHs and associated satellites}
\label{sec:observations}
On the basis of the accretion rates computed in the previous section
we now determine the corresponding bolometric accretion
luminosities. In the absence of any detailed spectral modelling, these
provide at least an indication of the magnitude of expected observational
signatures.

The  bolometric luminosity $L_{bol}$ is directly proportional to the
physical mass accretion
rate times the radiative efficiency parameter $\eta$
\begin{eqnarray} \label{eq:Lbol}
	L_{bol} &=& \eta \frac{dM}{dt} c^2 \nonumber \\
&=& \eta ~5\times 10^{35} \left(\frac{M_{\bh}}{100
M_{\odot}}\right)^2 \left(\frac{\rho_g}{10^{-24} {\rm ~g
~cm}^{-3}}\right) \nonumber \\
& & \times \left(\frac{c_s}{10 {\rm ~km ~s}^{-1}}\right)^{-3} (1 +
\beta_s^2)^{-3/2} {\rm ~erg ~s}^{-1}
\end{eqnarray}
and $c$ is the speed of light.
$\eta$ can reach maximum values of $\eta_{max} \approx 0.06$ for
non-rotating black holes and up to $\eta_{max} \approx 0.4$ for maximally rotating black
holes \cite{shapiro83}, if a mechanism for the effective
dissipation of energy exists. This mechanism is provided for by the
viscosity of matter in the accretion flow.
\subsection{Bolometric luminosity for accretion from the host ISM}
The effect of ISM accretion and the resulting emission has previously been
investigated analytically for stellar mass BHs in our galaxy that are the
remnants of ordinary stellar evolution \cite{fujita98}.
For MBHs larger luminosities are expected, particularly when they travel
within or are crossing the galactic disk and bulge regions or in
molecular clouds, as more gas is available. But even in less
dense regions of the inter stellar medium (ISM) MBHs could generate
sizable luminosities if they travel at low velocities and have
a correspondingly larger accretion radius.

In particular ISM turbulence can establish a geometrically thin  accretion disk around
the MBH with an associated radiative efficiency that can reach the
maximum values mentioned above.
For our computations we have adopted a `standard' thin disk radiative efficiency 
of $\eta_{td} = 0.1$, and used the ISM accretion rates determined in
the last section. 

\begin{figure*}{}
  \includegraphics[width=8.5cm]{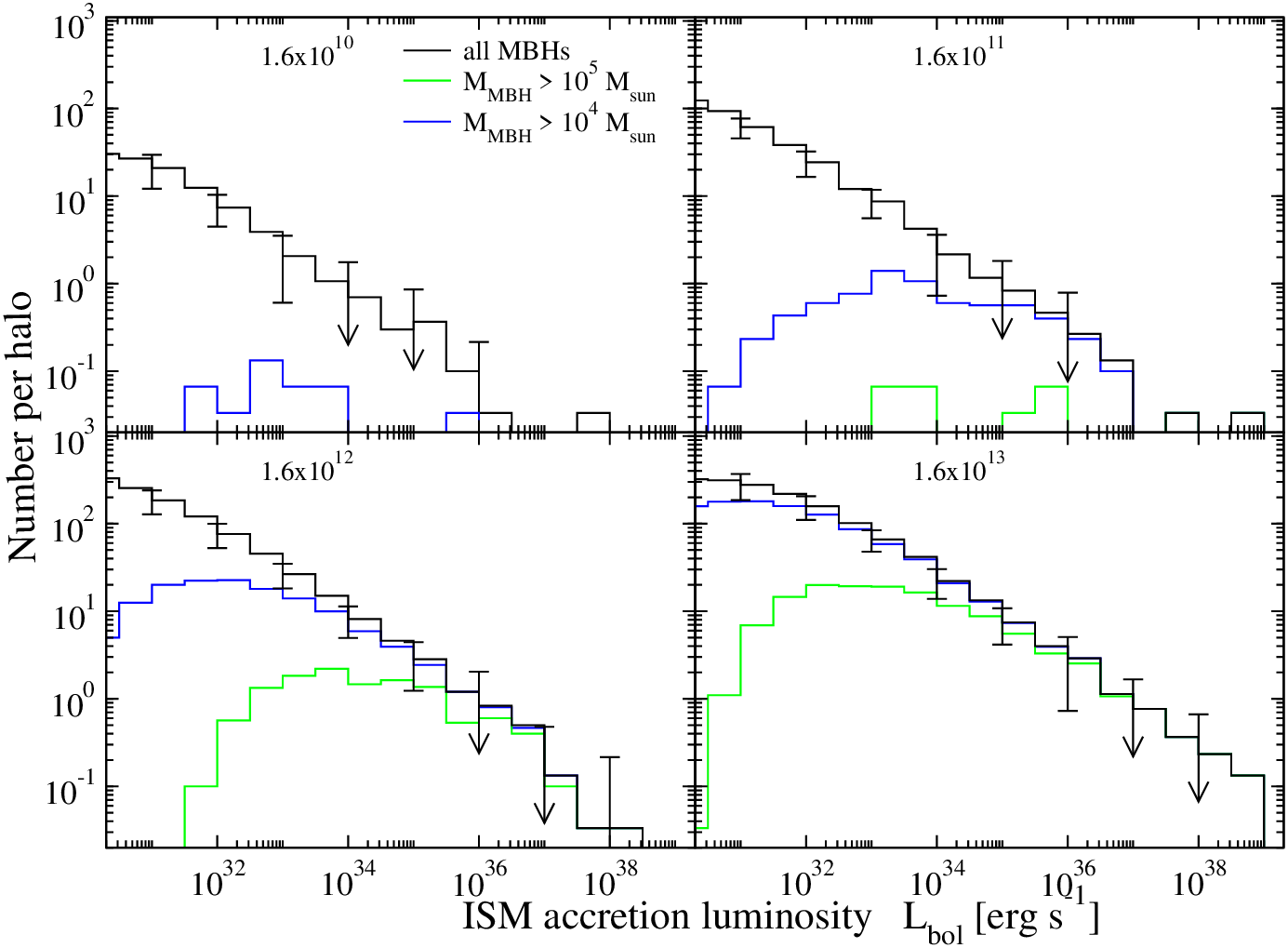}
  \vspace{0.5cm}
  \includegraphics[width=8.5cm]{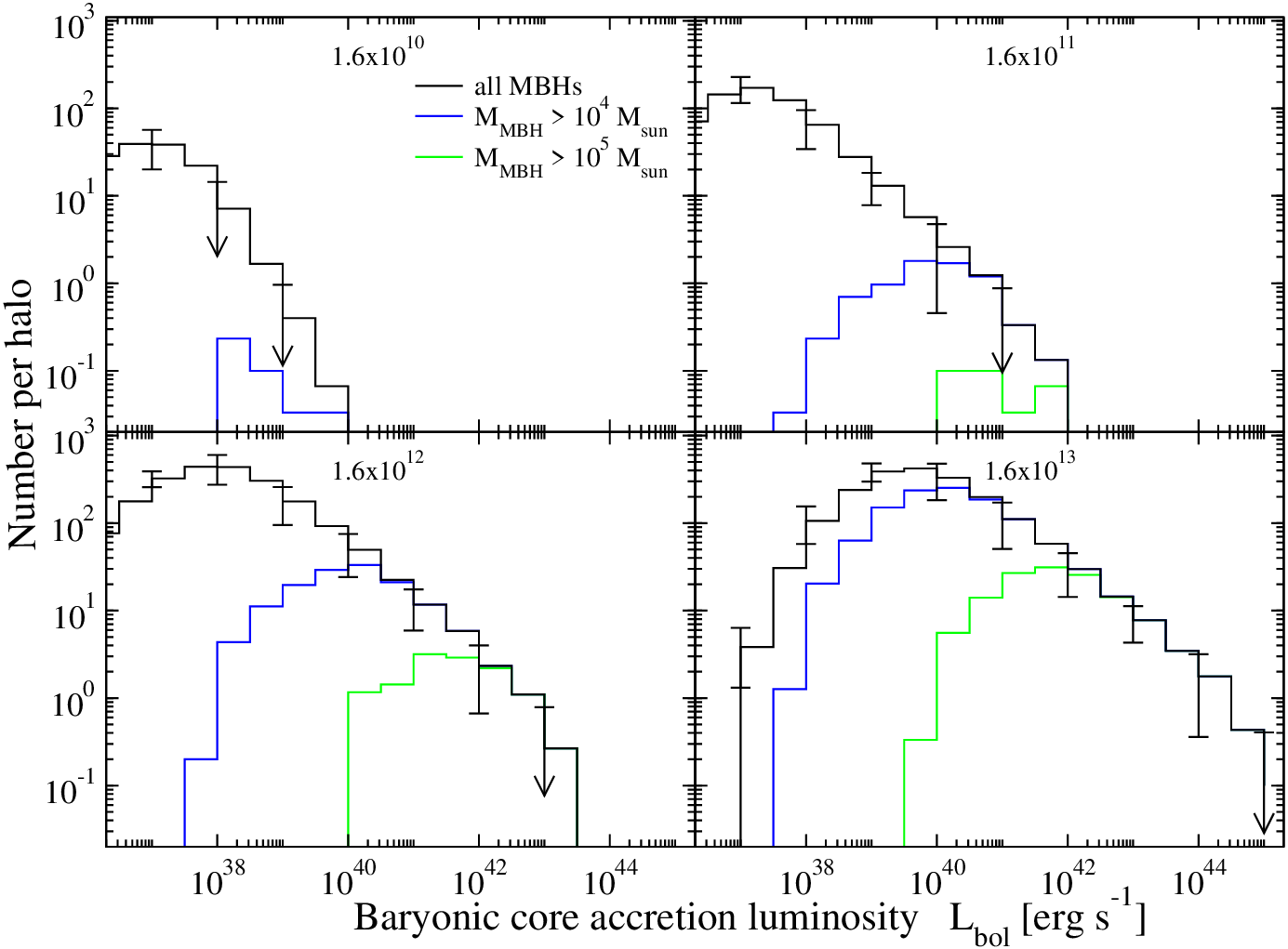}
  \caption{Bolometric luminosity functions for model C in the case of
    accretion from the host ISM (top panel) and from baryonic cores
    (bottom panel). Results are shown for all final halo
    masses and MBH mass cuts as shown.}
  \label{fig:Lbol}
\end{figure*}

The resulting bolometric luminosity function for the MBHs in
individual haloes is shown in the top panel of figure \ref{fig:Lbol}. This also shows
the contribution of MBHs above various mass thresholds.

In the last section we have seen that ISM accretion occurs at a very
much smaller rate than that from baryonic cores and the question is
how low the resulting accretion luminosity would be.
In fact from figure \ref{fig:Lbol} it becomes clear that MBHs accreting from the host ISM
have luminosities so low that it appears very difficult to detect them
let alone identify them as MBHs.
It might still be possible to derive
statistical constraints were the number of objects large enough
\footnote{See e.g. Fujita et al. (1998) who consider of the order of
$10^6$ BHs kpc $^{-3}$  \cite{fujita98}}.
For instance in an inhomogeneous ISM, that we have not accounted for
in our model, BHs travelling through dense regions in the 
galactic disk or bulge would emit at significantly larger
luminosities. A MBH accreting within a dense cloud that has a
density a factor 10 higher than the average, say, would see its
luminosity boosted by about the same factor
(c.f. eq. \ref{eq:Lbol}). By conservation of mass, however, these
clouds would only fill $ \lesssim 1/10$ of the ISM volume. If we have
10 MBHs uniformly distributed across the host ISM and accreting at
$10^{36} {\rm erg ~ s}^{-1}$ only at most one would end up in a cloud
with ten times larger density at any one time.
However, in our case the numbers of MBHs are just too low for
this process to be significant.

Figure \ref{fig:Lbol} shows the result for model C, which serves to
illustrate that even for the model that produces the highest  number of
large MBHs in the most massive haloes, the number of MBHs
accreting at bolometric luminosities larger than $\sim 10^{37} {\rm erg
  ~s}^{-1}$ is insignificant.
The main cause for these very low ISM accretion rates is that most
MBHs orbit at distances larger than the light radius
of the galaxy, which means that even if accretion luminosities were
very large in the disk and bulge we would still only observe very few
sources there.

Regarding this last point, baryonic core accretion also offers the advantage that it is essentially
independent of the structure and geometry of the host ISM.
\subsection{Bolometric luminosity for accretion from baryonic core remnants}
\label{sec:baryonic_cores}
\begin{figure*}
  \includegraphics[width=12cm,height=15cm]{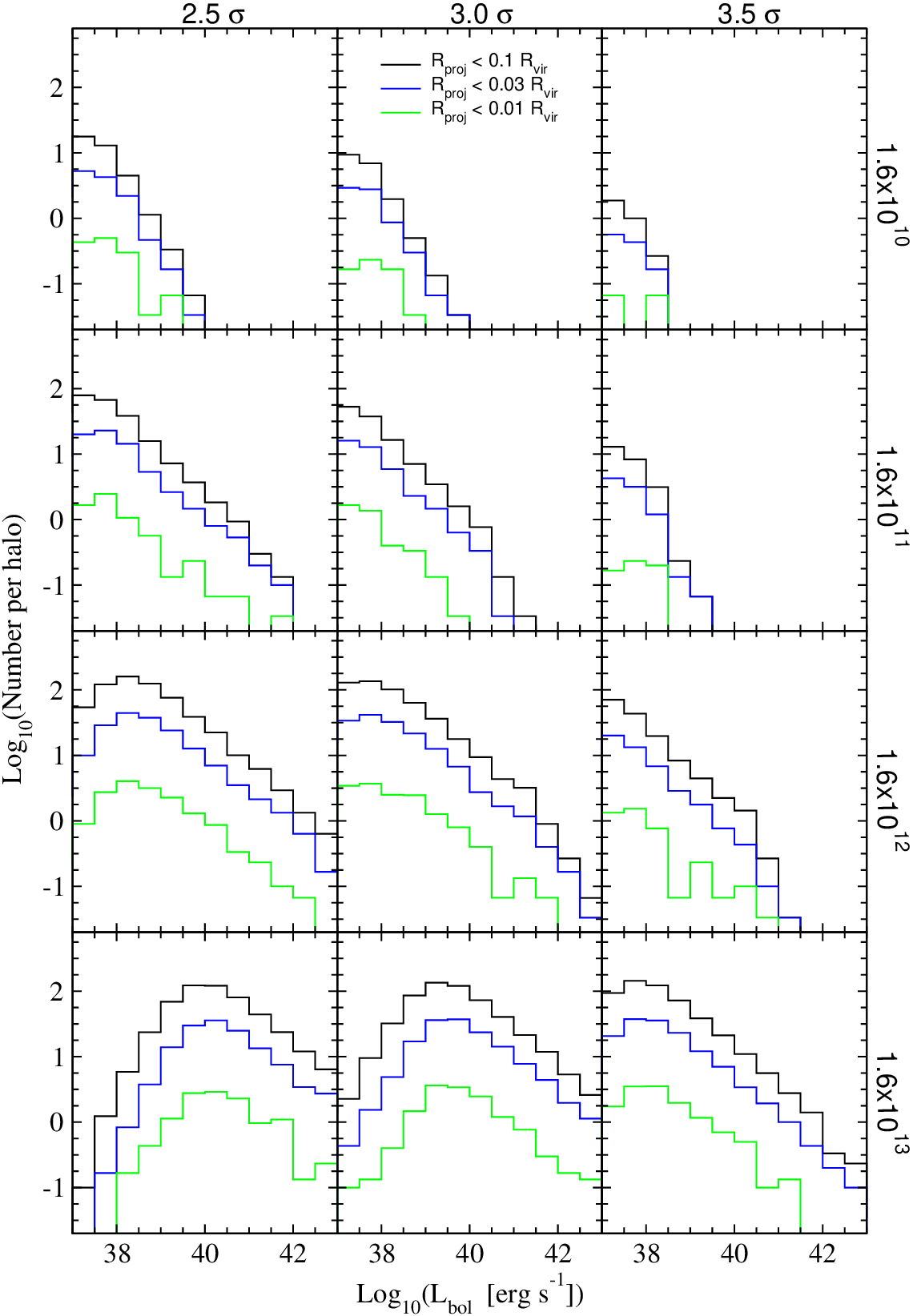}
  \caption{Bolometric luminosity of MBHs accreting from baryonic
    cores for models C, A and D (left to right panels). Shown are the
    sources whose line of sight falls within some
    projected distance from the host centre.}
  \label{fig:Lbol_project}
\end{figure*}
In the previous section we argued that MBHs accrete from a disk
assuming that the net specific angular momentum created in the
surrounding host ISM through turbulence is large enough.

For accretion from baryonic cores the required net angular momentum
comes for the largest part from the angular
momentum of the original satellite. Although the outer parts of
satellites may have been stripped and their angular momentum deposited in the 
host halo, the cores will preserve most of their angular momentum.
In this sense we can view these MBH - baryonic core systems as the
engines of mini-AGN stripped of their halo/galaxy within which they
originally resided.

Based on the baryonic mass accretion rates obtained in the previous
section and using the 10 \% standard thin disk efficieny, we obtain a
differential MBH bolometric luminosity function 
which is shown in the right panel of
figure \ref{fig:Lbol} for model C. $\beta_s = 0$ in this case as the MBHs
have no relative motion with respect to the baryonic cores from which
they accrete.
Figure \ref{fig:Lbol_project} shows what we call the projected
bolometric luminosity function, that is of all sources within a halo,
whose lines-of-sight (LOS) fall within some distance from the centre normal
to the LOS. 
Overall we see that the accretion luminosities are much higher than
for the ISM accretion case. They attain values larger than $10^{40} {\rm
  erg ~s}^{-1}$, although luminosities are somewhat lower for
models A and D with their lower MBH mass density.

The luminosity functions for the two most massive haloes exhibit a
logarithmic slope in the declining part of the function.
\begin{equation}
	N_{\bh} \propto L_{bol}^{-0.6 \pm0.02}
\end{equation}
Also from equation \ref{eq:accrate_vs_MBHmass}
\begin{equation}
  L_{bol} \propto \dot{M} \propto M_{\bh}^{1.68 \pm 0.02}
\end{equation}
Comparing with equation \ref{eq:Lbol} this implies a non-trivial scaling 
relation between $M_{\bh}$ and the satellite gas density
\begin{equation}
	M_{\bh} \propto \rho_g^{3.13 \pm 0.2} 
\end{equation}
Across all luminosities the largest fraction of the sources
is distributed throughout the host. Many of the brightest ones are in fact 
at distances larger than 10 percent of the host virial radius
$R_{vir}$.
The numbers of sources in the bulge and the disk are similar and
relatively low. Inside our galaxy most of the sources should therefore not be affected
by strong absorption in the disk and bulge. For other galaxies,
however, these problems remain for those MBHs whose line-of-sight
passes close to the central region of the galaxy within which they are 
located.
\subsection{Dark matter spikes and annihilations}
Accretion plays an important role in SMBH growth, as inferred  both from the
preceding discussion of the SMBH-spheroid correlation and
independently from the coincidence between black hole mass density and
the integrated, accretion-fed,  luminosity density of quasars.
If the MBH
grew  adiabatically, for example via gas accretion onto a seed MBH,
a spike develops in  the CDM density profile within the   region 
of gravitational influence of the MBH, $R\sim GM_{MBH}/\sigma^2,$ where 
$\sigma$ is the minihalo central velocity dispersion.
The cusp profile steepens from $\rho\propto r^{-\gamma}$ to
$\rho\propto
 r^{-\gamma^\prime}$, with 
$\gamma^\prime=\frac{9-2\gamma}{4-\gamma}$ \cite{gondolo99}.
Of course the  central SMBH  did form  by mergers, and  such a spike
would have been disrupted \cite{ullio01}. 
However most of the surrounding MBHs have not suffered a recent
merger,
within the age of the spheroid, and the CDM spikes would have survived
or been renewed as the MBH grew by accretion. 

It is generally believed that  the CDM consists of
neutralinos, the favoured stable massive MSSM (Minimal Super Symmetric
Model) candidate whose relic density was
determined by the annihilation rate in the early universe at the epoch
of thermal freeze-out. These neutralinos continue to annihilate today 
in the dark
halos, albeit at very slow rates that in principle can be calculated
by spanning MSSM parameter space once the relic abundance is
specified.
The CDM spikes result in a greatly enhanced annihilation rate
that can yield potentially detectable byproducts such as high energy
gamma rays, positrons and antiprotons. Indeed it has even been
suggested that an unidentified EGRET source in the direction of the
galactic centre might be   due to  such a spike  \cite{bertone02}.
One motivation for this suggestion was that the EGRET spectrum 
is too hard to be consistent with hadronic interactions by cosmic rays
and is consistent with an annihilation origin.

However, a reanalysis of
the EGRET data has led to a new positional identification of the GC
EGRET source not with  SagA* but with a nearby massive star cluster
(the Arches) \cite{hooper02}. Also, the unexpected presence of massive stars with
plunging orbits in the vicinity of the SMBH
associated with  SagA* \cite{ghez03} has been attributed to  the infall of 
massive stars under the gravitational influence of a IMBH (a MBH of
mass $10^3 - 10^4$ \Msun) that
represents the robust core of a star cluster able to survive tidal
disruption  and end up in the vicinity of  SagA* \cite{hansen03}.
We conjecture that this IMBH and others are relics of disrupted
minihalos and their baryonic cores, which still contain  CDM spikes.
One might therefore be seeing off-centre gamma ray sources
associated with spike annihilations. Indeed other unidentified EGRET
sources might also be due to relic annihilation spikes. The predicted 
$gamma$-ray flux is uncertain by several orders of magnitude due to
uncertainty in the the initial CDM cusp profile prior to spike
formation and to the MSSM parameter space. One could therefore 
be seeing MBHs at 10 kpc distance of mass well below $10^6$ M$_\odot$,
and possibly down to $10^3$ M$_\odot.$ Predicted spectra should be
hard, possibly extending to several  100 GeV, since the expected neutralino mass
range is approximately 50 GeV--2TeV.

\section{Summary and Conclusions}
\label{sec:conclusions}
We have used a semi-analytical approach to track the merger
history of massive black holes and their associated dark matter haloes,
as well as the subsequent dynamical evolution of the MBHs within the
new merged halo.
In particular we have looked at the possibility that MBHs that are the 
remnants of massive population III stars, forming in low mass haloes at 
redshifts $z \sim 20 - 30$, could hierarchically build up to contribute to
the present-day abundance of central galactic SMBHs.
If this is the case then a number of remnant MBHs are expected to orbit 
inside galactic haloes.
We expect our results to be representative for $\Lambda$CDM
cosmological models.

Our results can be summarised as follows:
\begin{itemize}
  \item For a range of models that consider different seed MBH masses as well as
    various collapse thresholds and redshifts for minihaloes we have
    determined the abundance of remnant MBHs in present day galactic
    haloes.
    As a result we expect of the order of a 1000 MBHs mostly with masses near
    the initial seed mass at the lower end and a a few as massive as
    $10^4$ to $10^5$ \Msun.  
  \item We have considered two accretion scenarios for the MBHs. For
  the case that MBHs accrete from the host ISM we found that the
  resulting accretion rates are too small for any related
  observational signatures to be significant. Instead accretion
  from baryonic core remnants of the satellite haloes that MBHs are/were
  originally associated with yields much larger accretion rates.
  Assuming a radiative efficiency of 10\%, a few MBHs within the
  visible extent of a galaxy
  would be expected to display bolometric accretion luminosities
  in excess of $10^{41} {\rm erg ~s}^{-1}$. This appears inconsistent with
  observations of ultraluminous off-centre  X-ray sources that have been
  detected in a number of galaxies (see e.g. Colbert \&
  Mushotzky 1999 \nocite{colbert99}). However, since most MBHs in our
  model accrete at
  only a fraction of the Eddington rate, this raises the possibility of
  radiatively inefficient accretion flows with correspondingly lower
  accretion luminosities. This option will be explored further in a
  subsequent paper.
  \item The slope of the $M_{SMBH} - M_{bulge}$ relationship in our
  model almost exactly matches that of observations, perhaps indicating
  that MBH merging really is a generic part of SMBH growth. Depending on the
  formation redshift of minihaloes and assumed mass of seed MBHs,
  however, various amounts of gas accretion are
  required to also match the normalisation of the observed
  relation. 
  \item Our model complements gas accretion based growth models for
  MBHs. It produces an
  appropriate number of MBHs and SMBH at high redshifts without which
  gas accretion alone could not explain the most massive SMBHs today as
  well as the presence of QSOs at redshifts of 6.
\end{itemize}

Our numerical results depend on a number of parameters that are not
yet well constrained, notably the primordial initial mass function of metal poor
stars forming inside minihaloes. However, we have shown
that, particularly for the abundance of MBHs in the halo, our results
hold qualitatively for a wide range of different primordial IMFs,
provided stars turn into MBHs of similar mass.

If the halo MBHs could be uniquely identified by their X-ray emission
or otherwise, then within the context of our model they could also be used
to tag (remnants of) substructure orbiting in a galactic halo. In this 
way they would complement counts and location  of dwarfs and star
clusters as measures of substructure in the galaxy and the halo.

Our results for the growth and present-day mass of the central SMBHs do 
depend sensitively on how efficiently MBHs merge at the host
centre. Here we have taken the view that during major mergers any MBHs 
orbiting within the core region of the host will be quickly dragged towards the 
central SMBH, aided by the massive inflow of gas. Inspiral of
the MBH by dynamical friction could further be boosted by the presence 
of a high density baryonic and dark matter core that remains associated
with the MBH and could thus potentially increase the mass by orders of
magnitude.

\section*{Acknowledgments}
The authors wish to thank R. Bandyopadhyay, G. Bryan, J.
Magorrian and H.-W. Rix for helpful discussions.
RRI acknowledges support from Oxford University and St Cross College, Oxford.
JET acknowledges support from the Leverhulme Trust.


\end{document}